\documentclass[prb,twocolumn,english,aps,prb,showkeys,a4paper,floatfix]{revtex4}
\usepackage[T1]{fontenc}
\usepackage{tikz}
\usepackage{color}
\usepackage{amsmath}
\usepackage{amssymb}
\usepackage[latin1]{inputenc}
\usepackage{graphicx}
\usepackage{bm}
\usepackage{epsfig}
\usepackage{xcolor}
\definecolor{mynicegreen}{RGB}{102,182,102}

\begin{document}

\title{Orientational ordering in a fluid of hard kites: A density-functional-theory study}

\author{Yuri Mart\'{\i}nez-Rat\'on}
\email{yuri@math.uc3m.es}
\affiliation{
Grupo Interdisciplinar de Sistemas Complejos (GISC), Departamento
de Matem\'aticas, Escuela Polit\'ecnica Superior, Universidad Carlos III de Madrid,
Avenida de la Universidad 30, E-28911, Legan\'es, Madrid, Spain}

\author{Enrique Velasco}
\email{enrique.velasco@uam.es}
\affiliation{Departamento de F\'{\i}sica Te\'orica de la Materia Condensada,
Instituto de F\'{\i}sica de la Materia Condensada (IFIMAC) and Instituto de Ciencia de
Materiales Nicol\'as Cabrera,
Universidad Aut\'onoma de Madrid,
E-28049, Madrid, Spain}

\date{\today}

\begin{abstract}
	Using Density Functional Theory we theoretically study the orientational properties of  
	uniform phases of hard kites -- two isosceles triangles joined by their common base. Two approximations are used: Scaled Particle Theory, 
	and a new approach which better approximates third virial coefficients 
	of two-dimensional hard particles. By varying some of their geometrical parameters kites can be transformed into 
	squares, rhombuses, triangles, and also very elongated particles, even reaching the hard-needle limit. Thus 
	a fluid of hard kites, depending on the particle shape, can stabilize isotropic, nematic, tetratic and  
	triatic phases. Different phase diagrams are calculated, including those of rhombuses, and 
	kites with two of their equal interior angles fixed to $90^{\circ}$, $60^{\circ}$ and $75^{\circ}$. 
	Kites with one of their unequal angles fixed to $72^{\circ}$, which have been recently studied via Monte Carlo simulations, are also 
considered.	
	We find that rhombuses and kites with two equal right angles and not too large anisometry 
	stabilise the tetratic phase
	but the latter stabilize it to a much higher degree. By contrast, kites 
	with two equal interior angles fixed to $60^{\circ}$ stabilize the triatic phase to some extent, although 
	it is very sensitive to changes in particle geometry.
	Kites with the two equal interior angles fixed to $75^{\circ}$ have a phase diagram with both tetratic and 
	triatic phases, but we show the nonexistence of a particle shape for which both phases are stable at different densities. 
	Finally the success of the new theory in the description of orientational order in kites is shown by comparing with Monte Carlo simulations
	for the case where one of the unequal angles is fixed to $72^{\circ}$. These particles also present phase diagrams with stable 
	tetratic and triatic phases.
	
\end{abstract}


\maketitle

\section{Introduction}

The study of entropically-driven phase transitions in liquid crystals has been an active 
field of research from  the pioneering work of Onsager \cite{onsager}, having a major boost in the 80's and 90's
\cite{frenkel,veerman,allen,jackson,bolhuis,mulder} 
and continuing as a very active topic of research up to the present
\cite{kooij,wensink,cheng,schiling,cinacchi,odriozola,
yang,cuetos,dijkstra2,dijkstra3,rafael}. 
These works theoretically and experimentally 
showed that liquid-crystalline uniform phases, such as uniaxial or biaxial nematics (N), and
non-uniform phases such as smectic and columnar phases, can be stabilized solely by extremely 
short-ranged repulsive particle interactions. Several 
statistical-mechanical models were developed for the description of 
thermodynamic and structural properties of hard-body fluids in which the Helmholtz free-energy 
has only an entropic contribution, with Density Functional Theory (DFT) being one of the most 
successful theoretical tool in this respect \cite{mederos}.

Most theoretical works naturally concentrated on 3D systems since in experiments  
the ratios between the lengths of the samples along the three spatial directions  
and those of the particles are large enough to conform to the three-dimensional spatial criterion. However new 
experimental techniques have been recently developed for the synthesis of taylor-shaped 
hard-core interacting microparticles, which can now be studied under extreme confinement along 
one spatial direction \cite{chaikin,zhao3,zhao5}. These systems can be thought of as single monolayers of particles 
subject to Brownian motion in two dimensions (2D). 

Recent experimental works 
on these effectively 2D hard-body fluids showed the stability of  
exotic uniform liquid-crystalline phases such as tetratic (T) \cite{chaikin,zhao3}, and triatic (TR) \cite{zhao5}. 
The symmetries of these phases can be rationalized from the properties 
of the orientational distribution function, $h(\phi)$, defined as the probability 
density for the angle $\phi$ between the particle axis and the nematic director. Four- or six-fold symmetries, i.e. 
$h(\phi)=h(\phi+2\pi/n)$, indicate the presence of T ($n=4$) or TR ($n=6$) phases, respectively.
Theoretical studies using MC simulations \cite{frenkel1,donev} 
and DFT \cite{schlacken,MR3,MR4} predicted the stability of the T phase long before the experiments were 
conducted. By contrast, theoretical studies of the TR phase \cite{dijkstra,MR2} appeared after 
the phase
was discovered in experiments \cite{zhao5}. 

Depending on their particular (usually polygonal) shape, 2D particles also crystallize 
into a variety of structures with different symmetries. These symmetries exhibit a subtle dependence 
on geometrical details such as the roundness of the particle corners. For example, in the case of regular 
polygons with more than seven sides, the crystal melts continuously into an hexatic phase and then transforms 
into an isotropic (I) fluid through a first order transition \cite{glotzer}. Triangles, squares and hexagons 
exhibit a Kosterlitz-Thouless transition from I to TR, T and hexatic phases, respectively, which then
crystallize \cite{glotzer}. Finally, pentagons undergo a one-step first-order 
melting from crystal to I \cite{glotzer}. However in a fluid of squares with rounded corners, 
the orientationally disordered hexatic-rotator, or orientationally-ordered rhombic crystalline phases are 
stabilized as density is increased \cite{zhao3,escobedo}, but no T phase was found; instead, an hexatic phase between 
I and crystal appears for a certain roundness parameter \cite{escobedo}. 
In the case of hexagons with rounded corners a transition occurs between an hexagonal 
rotator crystal and an hexagonal crystal \cite{zhao4}. DFT studies 
revealed that non-polygonal particles such as 
hard rectangles \cite{schlacken,MR3,MR4} or superellipses close enough to the rectangular 
shape \cite{szabi} can stabilize the T phase when the aspect ratio is below 
a certain critical value. 

Some recent experimental works have shown the tendency of some achiral 2D 
particles, such as equilateral triangles or square crosses,  
to form chiral crystalline structures at high packing fractions \cite{zhao5,mayoral,zhao6}. 
By mixing particles with exotic geometries, e.g. kites and darts, it
is also possible to obtain quasi-periodic structures in which kite- and dart-shaped tiles 
form pentagonal stars, arranged in turn into different close-packed superstructural patterns \cite{zhao7}. 
In recent experiments the phase diagram of kites with one of its unequal interior angles, $\alpha_1$, 
fixed to $72^{\circ}$ with the other, $\alpha_2$ being variable, was 
elucidated \cite{zhao1}. Interestingly, kites with a shape departing from the 
square geometry also form a T phase for some values of $\alpha_2$ \cite{zhao1}. 

In this work the phase behavior 
and orientational properties of a uniform fluid of hard kites is studied theoretically.
DFT is used, based on two alternative approximations: the standard Scaled Particle Theory (SPT),
and a new approach which better approximates the third virial coefficient.
Different constraints on the interior angles $\alpha_i$ of kites are selected.
The kite geometry has the square ($\alpha_i=90^{\circ}$) and equilateral triangle 
($\alpha_1=60^{\circ}$, $\alpha_2=180^{\circ}$)
as limiting cases. Actually, these shapes maximize T and TR stability, respectively. 
We are interested in changes in the stability of these phases resulting from distortions 
of these two polygonal geometries, always within the kite-like shape. Therefore, the following 
constraints on the interior angles are applied: (i) $\alpha_1=\alpha_2$ (rhombuses 
having the square as a limiting case), (ii) $\alpha_1+\alpha_2=180^{\circ}$  (kites with the same 
limiting case and with both equal interior angles fixed to $90^{\circ}$), (iii) 
$\alpha_1+\alpha_2=240^{\circ}$ (kites with the equilateral triangle as a limiting case), and 
(iv) $\alpha_1+\alpha_2=210^{\circ}$ (kites with the limiting case consisting of an isosceles triangle with opening angle 
equal to $30^{\circ}$). Phase diagrams of all these cases are calculated.
By comparing the first and second cases we show that the latter has a larger stability region for the T phase, 
i.e. a larger interval for $\alpha_1$ where the phase is stable. Also, the TR phase of hard equilateral triangles 
is very sensitive to changes in particle geometry, resulting in the lowest $\alpha_1$-interval for which the
phase is stable. The case (iv) is very interesting since the phase diagram presents stable T and TR phases for 
different $\alpha_1$. The existence of a particular particle shape 
(with fixed $\alpha_i$) that exhibits both phases at different densities can be discarded. 
Finally, we calculated the phase diagram of kites with one 
of the unequal angles, $\alpha_1$, fixed to $72^{\circ}$, while the other one, $\alpha_2$,
is freely varied. This study allowed our new DFT theory to be contrasted with the recent Monte Carlo (MC) simulations 
of Ref. \cite{zhao1}. We showed that, by varying $\alpha_1$ in the interval 
$[54^{\circ},180^{\circ}]$, both phases, T and TR, are stable, with the former having the largest stability region.
The size of this region is similar to that found in the simulations. This result, together with the agreement in the
values of packing fractions at the I-T transition, gives support to the validity of our DFT approach.

\section{Theory}
\label{theory}
In this section we introduce the theoretical tools used to study the equilibrium properties 
of the fluid of hard kites. In Sec. \ref{dft} the two version of DFT used are presented: the one based on the classical
SPT, and a new one which better approximates three-body correlations in general systems of 2D hard convex particles. 
In Sec. \ref{excluded_area} the particle model used and the properties of the excluded area (the main ingredient of the DFTs) are 
considered. Finally Sec. \ref{bifurcation} presents a bifurcation analysis using both theories 
to calculate  
the I-(TR,T) bifurcation curves; when the SPT approximation is used analytic expressions 
can be obtained. 
The bifurcation analysis from the (T,TR) phases to the N phase is described in Sec. \ref{appendix}.

\subsection{DFT for 2D hard convex particles}
\label{dft}

The density expansion of the fluid pressure is based on the knowledge 
of the virial coefficients $B_n$. For hard spheres or hard disks 
these coefficients are known to high order. However for anisotropic hard bodies only the cases $n=2$ and $3$ are available 
in general, and the latter case is only known for a few geometries.
In 2D the exact second-virial coefficient of convex bodies in the orientationally disordered I phase is given 
by \cite{kihara,tarjus}
\begin{eqnarray}
	B_2=a+\frac{{\cal L}^2}{4\pi}=a\left(1+\gamma\right),
	\label{second}
\end{eqnarray}
with $a$ and ${\cal L}$ the area and perimeter of the particle. The anisometry parameter 
\begin{eqnarray}
	\gamma=\frac{{\cal L}^2}{4\pi a},
\end{eqnarray}
is a measure of how much the particle geometry deviates from a disk. In this case $\gamma=1$, while for other convex particles 
$\gamma> 1$. 

A good approximation for the third virial coefficient, again for orientationally disordered particle configurations, 
is given by 
\begin{eqnarray}
	B_3=a^2+\delta_1 \frac{{\cal L}^2a}{4\pi}
	+\delta_2\frac{{\cal L}^4}{\left(4\pi\right)^2}
	=a^2\left(1+\delta_1\gamma+\delta_2\gamma^2\right),
	\label{b3}
\end{eqnarray}
where $\delta_i$ are numerical coefficients obtained by fitting the available values of $B_3$  
(calculated from MC integration) for several convex particles \cite{boublik,tarjus,MR4}.

An interesting limit is the Onsager hard-needle limit, where particles become infinitely 
elongated. In this limit the particle aspect ratio $\kappa$ becomes infinite, $\kappa\to\infty$.
The behavior of the ratio of $B_3$ to $B_2^2$ is \cite{onsager,tarjus}  
\begin{eqnarray}
	&&\lim_{\kappa\to\infty}\frac{B_3}{B_2^2}=0,\ \text{in}\ \text{3D},\label{3d}\\
	&&\lim_{\kappa\to\infty}\frac{B_3}{B_2^2}\simeq 0.514,\ \text{in}\ \text{2D} \label{2d}.
\end{eqnarray}
The 3D limit explains the success of DFT theories for 3D hard-body fluids based only on the 
exact second virial coefficient. By contrast, because of (\ref{2d}), the corresponding 2D theories have a lesser degree 
of accuracy and third and possibly higher-order virial coefficients are necessary in the theory to adequately 
account for particle correlations in the fluid.

For orientationally ordered phases, the anisometry parameter becomes a functional of the 
orientational distribution function $h(\phi)$:
\begin{eqnarray}
	\gamma[h]&=&\frac{\langle\langle A_{\rm spt}(\phi)\rangle\rangle_{h(\phi)}}{2a}\nonumber\\
	&\equiv& \frac{1}{2a}\int_0^{2\pi} d\phi_1\int_0^{2\pi} d\phi_2 h(\phi_1)h({\phi_2})
	A_{\rm spt}(\phi_{12}),
	\label{gamma}
\end{eqnarray}
This is defined as a double angular average of $A_{\rm spt}(\phi)$, which is directly related to the 
excluded area between two particles as
\begin{eqnarray}
	A_{\rm spt}(\phi)\equiv A_{\rm excl}(\phi)-2a.
\end{eqnarray}
Note that, inserting the uniform distribution function $h(\phi)=(2\pi)^{-1}$ in (\ref{gamma}), we obtain 
\begin{eqnarray}
	\gamma=\frac{1}{4\pi a}\int_0^{2\pi} d\phi A_{\rm spt}(\phi)=\frac{{\cal L}^2}{4\pi a}.
\end{eqnarray}
The latter equality is proven in Refs. \cite{isihara,kihara} for general convex particles. Following a similar 
reasoning, an approximation for the third-virial coefficient of orientationally 
ordered phases can be obtained by substituting the value of the anisometry parameter by 
its functional form $\gamma\to \gamma[h]$ in (\ref{b3}). 

For perfectly oriented nematic phase, with the symmetric orientational distribution function 
$h(\phi)=\left[\delta(\phi)+\delta(\phi-\pi)\right]/2$ ($\delta(x)$ being the Dirac-delta function),   
one obtains from Eq. (\ref{gamma}) $\gamma[h] = \left[A_{\rm excl}(0)+A_{\rm excl}(\pi)\right]/(4a)-1$. 
Now taking into account that the excluded area of perfectly antiparallel oriented 
convex particles is equal to four times the particle area, $A_{\rm excl}(\pi)=4a$, 
we obtain $\gamma[h]=A_{\rm excl}(0)/(4a)$. Finally if particles are symmetric 
($A_{\rm excl}(0)=A_{\rm excl}(\pi)$) the same value as for hard disks, $\gamma[h]=1$, is obtained.

Following the SPT approximation, the excess free-energy per particle in thermal units $kT=\beta^{-1}$ is given by 
\begin{eqnarray}
	\varphi_{\rm exc}[h]\equiv \frac{\beta {\cal F}_{\rm exc}[h]}{N}
	=-\log\left(1-\eta\right)+\frac{\gamma[h]\eta}{1-\eta},
	\label{spt}
\end{eqnarray}
with $k$ the Boltzmann constant, $T$ the temperature and $N$ the total number of particles.
${\cal F}_{\rm exc}[h]$ is the Helmholtz free-energy density functional.
The fluid packing fraction is $\eta=\rho a$ with $\rho$ the number density. Note that the 
density expansion of (\ref{spt}), up to second order, is 
\begin{eqnarray}
	\varphi_{\rm exc}[h]&\simeq& a\left(1+\gamma[h]\right) \rho+\frac{1}{2}\left(1+2\gamma[h]\right)a^2 \rho^2\nonumber\\
	&=&B_2[h]\rho+\frac{1}{2}B_3^{(\rm spt)}[h]\rho^2.
\end{eqnarray}
This gives the exact expression for the second virial coefficient given by (\ref{second}), and an approximate 
value for the third one as 
\begin{eqnarray}
	B_3^{(\rm spt)}[h]=\left(1+2\gamma[h]\right)a^2.
	\label{b3_spt}
\end{eqnarray}
We note that the third virial coefficient for the I phase, as obtained from SPT, 
gives the incorrect hard-needle limit    
\begin{eqnarray}
	\lim_{\kappa \to \infty} \frac{B_3^{(\rm spt)}}{B_2^2}=
	\lim_{\kappa \to \infty} \frac{1+2\gamma}{\left(1+\gamma\right)^2}=0,
\end{eqnarray}
since $\gamma\to \infty$ as $\kappa\to\infty$. Comparing Eqns. (\ref{b3}) and (\ref{b3_spt}) we conclude  
that, for hard disks ($\gamma=1$), both expressions coincide if and only if $\delta_1+\delta_2=2$.

To overcome the failure of SPT to describe the correct scaling behavior in the Onsager limit, we here propose 
a different expression for the excess free-energy which gives the exact value of $B_2$ and the 
approximation (\ref{b3}) for $B_3$, resulting in the correct scaling for $\kappa\to\infty$. 
We also require to recover the SPT expression for hard disks, so we choose the condition $\delta_1+\delta_2=2$. 
Finally we set $\displaystyle{\delta_2=\frac{1}{2}}$ so that the hard-needle limit,
$\displaystyle{\frac{B_3}{B_2^2}\to 0.514\approx \frac{1}{2}}$, is accurately approximated. 

With these constraints in mind our proposal is
\begin{eqnarray}
	&&\varphi_{\rm exc}[h]=-\log(1-\eta)+\frac{\gamma[h]\eta}{1-\eta}\nonumber\\
	&&\hspace{-1cm}+\gamma[h]\left(\gamma[h]-1\right)\left(\frac{1}{2}+r\eta\right)
	\left(\frac{\eta}{1-\eta}+\log(1-\eta)\right),
	\label{new}
\end{eqnarray}
where an extra term is included, proportional to $r\eta$, which only affects  
the expressions for the fourth and higher virial coefficients. The coefficient $r$ can be chosen to
accurately describe the packing fraction of the I-T transition of the hard square fluid (as obtained from MC simulations). 

The ideal part of the free-energy per particle, dropping the thermal area, is
\begin{eqnarray}
	\varphi_{\rm id}[h]&\equiv& \frac{\beta {\cal F}_{\rm id}[h]}{N}\nonumber\\
	&=&\log \eta-1+\int_0^{2\pi}d\phi h(\phi)\log\left(2\pi h(\phi)\right).
\end{eqnarray}
The scaled fluid pressure $\displaystyle{\beta pa=\eta^2\frac{\partial \varphi}{\partial \eta}}$ is calculated from the
the total free-energy per particle $\varphi[h]=\varphi_{\rm id}[h]+\varphi_{\rm exc}[h]$ as
\begin{eqnarray}
	\beta p a &=&\frac{\eta}{1-\eta}+\frac{\gamma[h]\eta^2}
	{(1-\eta)^2}+\gamma[h]\left(\gamma[h]-1\right)\eta^2\nonumber\\&\times&
	\left[\left(\frac{1}{2}+r\right)\frac{\eta}{(1-\eta)^2}+
	r\log(1-\eta)\right].
	\label{EOS}
\end{eqnarray}
As already mentioned the anisometry asymptotically behaves as $\gamma[h]\sim 1$ for very high orientational ordering. 
Thus, in the case $(\gamma[h]-1)/(1-\eta)^2\sim 0$, Eqns. (\ref{new}) and (\ref{EOS}) show that the SPT 
(the first two terms in both equations) is also recovered at high packing fractions. 

In Sec. \ref{results} we use the SPT approximation (\ref{spt}) and our new proposal (\ref{new}) to calculate 
the phase diagrams of hard kites. As usual, the total free-energy per particle 
$\varphi[h]$ is minimised with respect to $h(\phi)$ to obtain its equilibrium value. 
The minimization is much less demanding numerically using truncated Fourier expansions for the orientational distribution function, 
\begin{eqnarray}
	h(\phi)=\frac{1}{2\pi}\left(1+\sum_{k=1}^n h_k \cos(2k\phi)\right),
	\label{fourier}
\end{eqnarray}
and then minimizing $\varphi[h]$ with respect to the Fourier coefficients $\{h_k\}$. The second order I-(T,TR) transition lines
are calculated using a bifurcation analysis (see Sec. \ref{bifurcation}), while the coexisting binodals are obtained 
from the equality of the chemical potentials $\displaystyle{\beta\mu=\varphi+\frac{\beta pa}{\eta}}$ 
and pressures $\beta p a$ (evaluated at the equilibrium values of $\{h_k\}$) in the two coexisting phases. 

\begin{figure}
	\epsfig{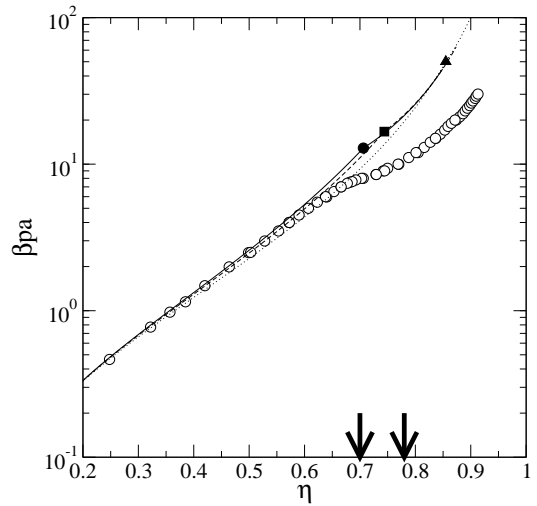}
	\caption{EOS of hard squares according to SPT (dotted), the new approach with $r=1$ 
	(dashed) and with $r=2$ (solid). Open circles represent MC simulation results from 
	Ref. \cite{frenkel1}. The I-T transitions from SPT and the new approach using $r=1$ and $r=2$ are shown 
	by solid triangles, squares and circles, respectively. The stability region of  
	the T phase obtained from MC simulations \cite{frenkel1} is defined by the arrows.}
	\label{fig1}
\end{figure}

The only uniform orientationally ordered phase in a fluid of hard squares is the tetratic 
phase. From the excluded area between two hard squares we obtain 
$A_{\rm spt}(\phi)=2a\left(|\sin\phi|+|\cos\phi|\right)$ (the key quantity to calculate $\gamma[h]$). 
The symmetry of the T phase implies $h(\phi)=h(\phi+\pi/2)$, and consequently
the Fourier expansion (\ref{fourier}) should only contain even integers $k=2j$ ($j\geq 1$). $\varphi[h]$ is then
minimised with respect to $\{h_{2j}\}$ for a given $\eta$, with $\varphi_{\rm ex}[h]$ given by SPT [Eq. (\ref{spt})], 
and also using our new proposal [Eq. (\ref{new})] with $r=1$ and $2$.  Inserting the equilibrium values 
into (\ref{EOS}) and its SPT-version (the first two terms), three different 
approximations for the equations of state (EOS) are obtained. Results are shown in Fig. \ref{fig1}, which also
includes the EOS of hard squares obtained from MC simulations, Ref. \cite{frenkel1}. 
The left arrow indicates the I-T transition from simulations, which occurs for $\eta\simeq 0.7$. 
The conclusion is that the choice $r=2$ predicts the transition much better, while the SPT gives a much higher 
value $\eta\approx 0.855$. The figure also shows how the theories  
overestimate the fluid pressure with respect to MC simulations, particularly close to the phase transition. It should 
be taken into account that simulation results also include the crystal phase 
at densities higher than $\eta\simeq 0.78$ (the right arrow in Fig. \ref{fig1}). 
However the crystal phase has not been included in our DFT study, so that 
it makes sense  that both theories overestimate the pressure at high densities.

The SPT approximation had been extensively used in the description of the phase behavior 
of hard particle fluids. As will be shown in Sec. \ref{bifurcation}, it has the advantage of producing analytic 
expressions of the packing fraction at the continuous transition from I  
to the orientationally ordered phases as a function of the particle characteristic lengths. Because of this 
we decided to calculate most phase diagrams with the SPT formalism. The new proposal (\ref{new}) 
was numerically implemented to calculate two different phase diagrams with the aim to comparing both theories. Also we 
wanted to confront the new theory with recent MC simulations for hard kites \cite{zhao1}. 

\subsection{Excluded area of hard kites}
\label{excluded_area}

\begin{figure}
	\epsfig{file=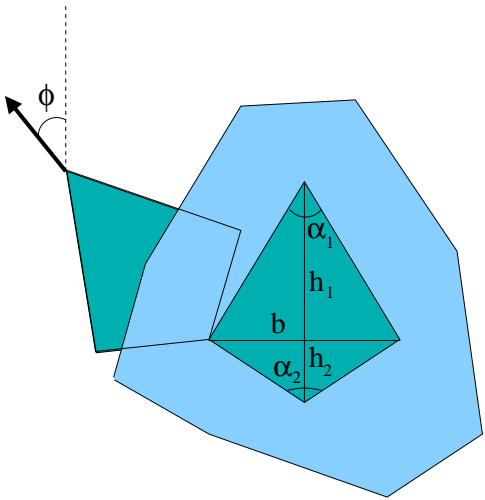,width=2.5in}
	\caption{Sketches of the particle geometry and the excluded area of two kites 
	for a relative angle $\phi$. Some characteristic lengths and 
	angles are shown.}
	\label{fig2}
\end{figure}

Kites are formed by two isosceles triangles of heights $h_1$ and $h_2$ and unequal opening 
angles $\alpha_1$ and $\alpha_2$ ($0\leq\alpha_i\leq\pi$), 
joined by their common bases $b$. See a sketch of 
the particle in Fig. \ref{fig2}. The other two interior angles, not indicated in the figure, are equal and have 
a value of $\pi-(\alpha_1+\alpha_2)/2$. In the same figure the excluded area between 
two kites with a relative angle $\phi$ is drawn. The particle axis is 
parallel to the heights and we choose the axis to point from the vertex with the largest opening angle
to that with the smallest one. The particle area is $\displaystyle{a=\frac{b}{2}(h_1+h_2)=
l_1l_2
\sin\left(\frac{\alpha_1+\alpha_2}{2}\right)}$ with $l_1$ and $l_2$ the lengths 
of the isosceles triangles, $l_i=\sqrt{h_i^2+b^2/4}$. 

Considering that $\alpha_1\leq \alpha_2$ (as sketched in Fig. \ref{fig2}), the SPT area, 
$A_{\rm spt}(\phi)=A_{\rm excl}(\phi)-2a$,
with a relative angle $0\leq \phi\leq \pi$ can be calculated from 
\begin{eqnarray}
	A_{\rm spt}(\phi)&=&2l_1^2\sin\alpha_1\cos\phi\nonumber\\&+&l_1^2
	\sin(\phi-\alpha_1)\Theta(\phi-\alpha_1)\nonumber\\&+&l_2^2\sin(\phi-\alpha_2)
	\Theta(\phi-\alpha_2)\nonumber\\
	&+&2l_1l_2\left[\sin(\phi-\alpha_{12}^-)
	\Theta(\phi-\alpha_{12}^-)\right.\nonumber\\&+&\left.\sin(\phi-\pi+
	\alpha_{12}^+)\Theta(\phi-\pi+\alpha_{12}^+)\right].
\end{eqnarray}
Here we have defined $\alpha_{12}^{\pm}=(\alpha_2\pm\alpha_1)/2$, and 
$\Theta(x)$ is the Heaviside function. For $\pi\leq\phi\leq 2\pi$ the 
SPT area is just $A_{\rm spt}(2\pi-\phi)$.

In general, kites are not symmetric with respect to $180^{\circ}$ rotations. However, as we showed in Ref. \cite{MR2}, 
a fluid of hard triangles (also a non-symmetric particle) has equilibrium N and TR phases  
with orientational distribution functions having the symmetry $h(\phi)=h(\pi-\phi)$, a property directly related to
the nonnegativity of the odd-index Fourier amplitudes of the function $A_{\rm spt}(\phi)$ \cite{MR2}:
\begin{eqnarray}
	\int_0^{2\pi}d\phi \cos[(2k-1)\phi] A_{\rm spt}(\phi)=\left\{
		\begin{matrix}
			0, & k=1,\\
			>0, & k>1.
		\end{matrix}
		\right.
		\end{eqnarray}
		The function $A_{\rm spt}(\phi)$ for kites also exhibits the same property.
		This symmetry of the  orientational distribution function 
		implies that particles axes have equal probabilities to point
		along the two possible directions parallel to the (N,T,TR)-directors.

\begin{figure}
	\epsfig{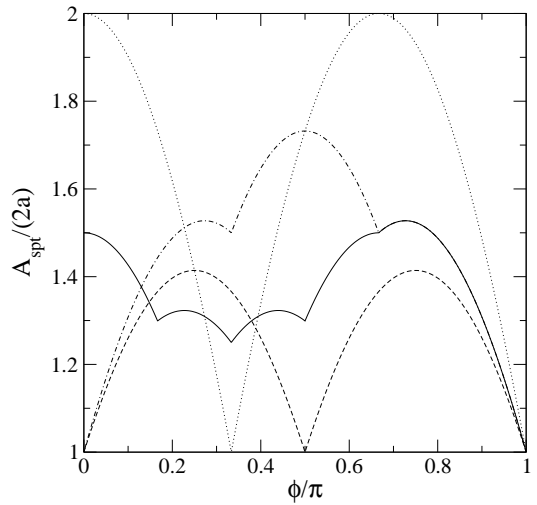}
	\caption{The function $A_{\rm spt}(\phi)$ for squares (dashed), triangles with 
	$\alpha_1=60^{\circ}$ and $\alpha_2=180^{\circ}$ (dotted), rhombuses with 
	$\alpha_1=\alpha_2=60^{\circ}$ (dot-dashed) and kites with $\alpha_1=60^{\circ}$ and 
	$\alpha_2=120^{\circ}$ (solid).}
	\label{fig3}
\end{figure}

By construction kites can degenerate into squares if $\alpha_1=\alpha_2=90^{\circ}$, 
into triangles when $\alpha_2=180^{\circ}$ and $\alpha_1<180^{\circ}$, or into rhombuses for $\alpha_1=\alpha_2$.
Fig. \ref{fig3} shows four examples of the function $A_{\rm spt}(\phi)$ for squares, 
equilateral triangles ($\alpha_1=60^{\circ}$ $\alpha_2=180^{\circ}$), 
rhombuses with $\alpha_1=\alpha_2=60^{\circ}$ and 
also for kites with  $\alpha_1=60^{\circ}$ and $\alpha_2=240^{\circ}$. The symmetries of 
this function are: (i) $A_{\rm spt}(\phi)=A_{\rm{ spt}}(\phi+\pi/2)$ for squares, (ii) 
$A_{\rm spt}(\phi)=A_{\rm{ spt}}(\phi+2\pi/3)$ for equilateral triangles, and (iii) 
$A_{\rm spt}(\phi)=A_{\rm{ spt}}(\pi-\phi)$ for rhombuses. These symmetries 
are directly related to the propensity of the system 
to stabilize the T, TR and N phases, respectively, at high densities.  
Also note the complexity of $A_{\rm spt}(\phi)$ for kites with 
$\alpha_1=60^{\circ}$ and $\alpha_2=120^{\circ}$ (this is generally true for $\alpha_1\neq\alpha_2$), 
with the presence of 
several local minima and maxima, and with the absolute minimum always located at $\phi=\pi$. 
Thus the minimum excluded area is always reached when the main particle axes are antiparallel,
resulting in $A_{\rm excl}(\pi)=4a$.

\subsection{Bifurcation analysis from I phase}
\label{bifurcation}
\begin{figure*}
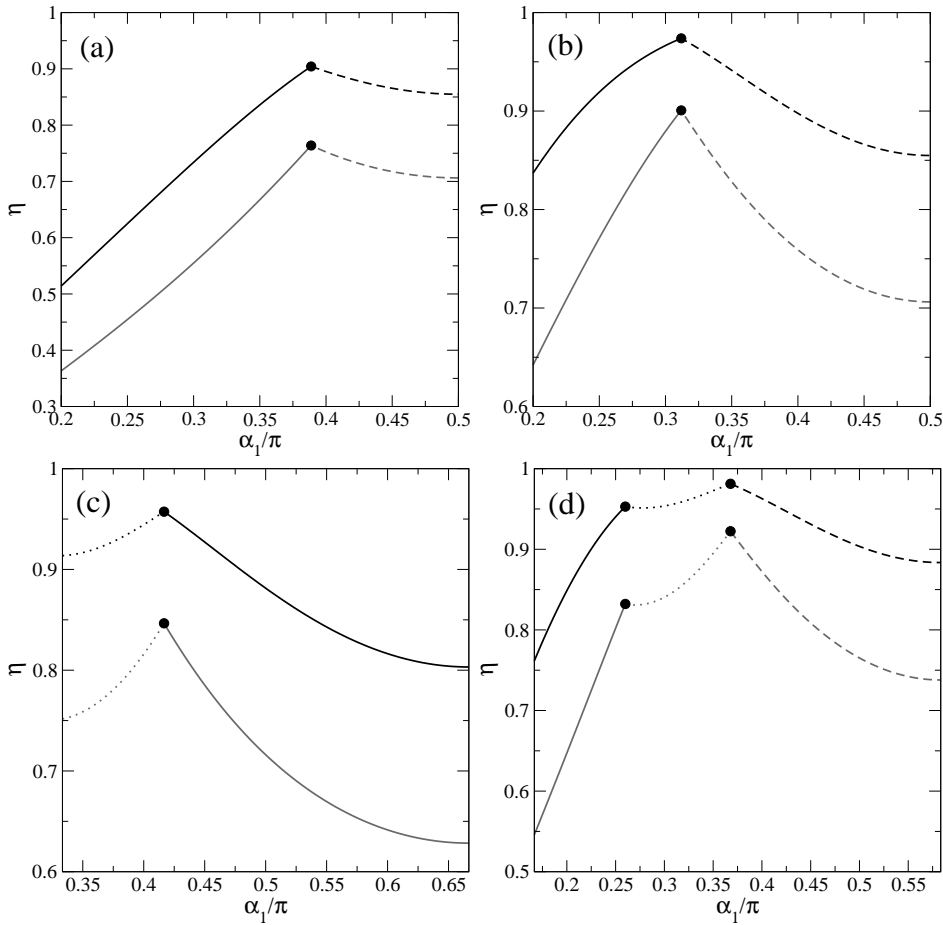

	\epsfig{file=fig4a.eps,width=2.4in}
	\epsfig{file=fig4b.eps,width=2.4in}
	\epsfig{file=fig4c.eps,width=2.4in}
	\epsfig{file=fig4d.eps,width=2.4in}
	\caption{I-N (solid), I-T (dashed) and I-TR (dotted) bifurcation curves 
	of (a) hard rhombuses, (b) kites with $\alpha_1+\alpha_2=180^{\circ}$, (c)
	$240^{\circ}=4\pi/3$ and (d) $210^{\circ}=7\pi/6$, according to SPT (black) and the new 
	approach (dark grey).
	Intersections between these curves are shown by black circles. Intervals of the opening angle $\alpha_1$ 
	shown in (c) and (d) are $[\pi/3,2\pi/3]$ and $[\pi/6,7\pi/12]$, respectively.}
	\label{fig4}
\end{figure*}

In this section we present the calculation of the second order transition lines, $\eta$ as 
a function of the opening angle $\alpha_1$ of kites, 
from the I to orientationally ordered phases, using some constraints on the other angle $\alpha_2$.
As shown in Sec. \ref{results} these transitions can be of first order. However 
this generally occurs in a small region of the phase diagram, so the I-(TR,T,N) transitions 
are, for most values of $\alpha_1$, of second order.

Inserting the Fourier expansion (\ref{fourier}) into the definition 
of $\gamma[h]$, Eq. (\ref{gamma}),  we obtain
\begin{eqnarray}
	\gamma[h]=\gamma_0+\frac{1}{2}\sum_{k\geq 1} \gamma_k h_k^2,
\end{eqnarray}
where we define the coefficients
\begin{eqnarray}
	&&\gamma_k\equiv 
	\frac{1}{\pi}\int_0^{\pi} d\phi \frac{A_{\rm spt}(\phi)}{2a}\cos(2k\phi)
	\nonumber\\
	&&\hspace{-1.0cm}=-\frac{\left[\sin(\alpha_1/2)\cos(k\alpha_2)+\sin(\alpha_2/2)\cos(k\alpha_1)
	\right]^2}{\left(4k^2-1\right)\pi\sin(\alpha_1/2)\sin(\alpha_2/2)\sin\left[
		(\alpha_1+\alpha_2)/2\right]}, \label{coeff}
\end{eqnarray}
depending only on $\alpha_i$ ($i=1,2$). Note that we used the symmetry of $A_{\rm spt}(\phi)$ 
with respect to the axis $\phi=\pi$ to integrate from 0 to $\pi$, multiplying the result 
by 2. In the following we use the same symmetry of $h(\phi)$ to change 
the integration intervals from $[0,2\pi]$ to $[0,\pi]$. Because of this, the normalization
factor $(2\pi)^{-1}$ in $h(\phi)$ (see Eq. (\ref{fourier})) will be substituted by $\pi^{-1}$.

Consider a small perturbation of the orientational distribution function of the I phase, 
$h(\phi)\approx \pi^{-1}\left(1+h_j^2\cos(2j\phi)\right)$, where $j=1,2$ and 3 
for N, T, and TR symmetries, respectively. The lowest order perturbation of the ideal 
part of the free-energy per particle is $\displaystyle{
	\varphi_{\rm id}\approx \log \eta-1+
\frac{h_j^2}{4}}$, while the excess part can be calculated from (\ref{new}), taking 
\begin{eqnarray}
	\gamma[h]\approx \gamma_0+\frac{1}{2}\gamma_j h_j^2,
\end{eqnarray}
and retaining only terms proportional to $h_j^{2n}$ (with $n=0,1$).
The free-energy difference between the orientationally ordered 
phase $\rm{X}$ ($\rm{X}=$N, T, TR) and the I phase is
\begin{eqnarray}
&&\hspace{-0.5cm}\Delta \varphi\equiv \varphi_{\rm X}-\varphi_{\rm I}\simeq 
	\left\{1+2\left[\frac{\eta}{1-\eta}+(2\gamma_0-1)\left(\frac{1}{2}+r\eta\right)\right.\right.\nonumber\\
	&&\left.\left.\hspace{1cm}\times\left(\frac{\eta}{1-\eta}+\log(1-\eta)\right)\right]\gamma_j\right\}
	\frac{h_j^2}{4}.
	\label{diff}
\end{eqnarray}
At the bifurcation point the factor inside the square bracket is equal to zero. The value of the 
packing fraction at this point is obtained by solving (\ref{diff})
numerically for $\eta$.
Considering now the free-energy difference from the SPT approach, i.e. the same Eqn. 
(\ref{diff}) but removing the term proportional to $2\gamma_0-1$, we obtain a simple 
analytical result:
\begin{eqnarray}
	\eta_j=\frac{1}{1-2\gamma_j},
\end{eqnarray}
where the packing fraction is labeled with $j$, indicating the 
symmetry of the bifurcated phase.
Some interesting cases are: rhombuses 
with $\alpha_1=\alpha_2$,  
and kites with $\alpha_1+\alpha_2=180^{\circ}$. The latter constraint implies that the 
other two equal angles of the kites are fixed to $90^{\circ}$. As shown below 
this restriction constitutes an important requirement for a stable T phase even 
for values of $\alpha_1$ significantly different from $90^{\circ}$ 
(square geometry). The expressions for $\eta_j$ for these important cases are 
\begin{eqnarray}
	\frac{1}{\eta_j}=\left\{
	\begin{array}{l}
			1+\frac{8\cos^2(j\alpha_1)}{\pi(4j^2-1)\sin\alpha_1}\hspace{2.5cm}
			(\alpha_1=\alpha_2)\\
			\\
			1+\frac{2\cos^2(j\alpha_1)\left[\tan(\alpha_1/2) 
			+\cot(\alpha_1/2)+2(-1)^j\right]}{\pi(4j^2-1)} \\
			\hspace{4.8cm}(\alpha_1+\alpha_2=\pi).
		\end{array}
		\right.\nonumber\\
		\label{bifurca1}
\end{eqnarray}
A first indication for the stability of the T phase in a fluid of 
hard rhombuses is given by the intersection of the I-N ($j=1$) and 
I-T ($j=2$) bifurcation curves, $\eta_1(\alpha_1)=\eta_2(\alpha_1)$. This equality
gives the result $\alpha_1^*\simeq 69.98^{\circ}$, a value corresponding to
the most anisometric rhombus with a stable T phase. In fact the actual value is a  
bit larger since, as shown in Sec. \ref{results}, the phase transitions are of first order
in the neighborhood of the intersection point. For the case $\alpha_1+\alpha_2=180^{\circ}$ 
the solution to the equation $\eta_1(\alpha_1)=\eta_2(\alpha_1)$ is 
$\alpha_1^*=56.14^{\circ}$. Obviously the fact that the other two equal angles of 
the kites are $90^{\circ}$ promotes the stabilization of the T phase for values of $\alpha_1$ lower to those 
for hard rhombuses.

Applying now the constraint $\alpha_1+\alpha_2=240^{\circ}$, we obtain the I-N and I-TR bifurcation 
curves
\begin{eqnarray}
	\frac{1}{\eta_j}=\left\{
		\begin{matrix}
	1+\frac{\cos^2(3\alpha_1/2)}{\sqrt{3}\pi\sin(\alpha_1/2)
			\sin(\alpha_1/2+\pi/3)} & (j=1) \\
			&\\
	1+\frac{4\sqrt{3}\cos^2(3\alpha_1)\cos^2(\alpha_1/2-\pi/3)}
			{35\pi \sin(\alpha_1/2)\sin(\alpha_1/2+\pi/3)} & (j=3).
		\end{matrix}
		\right.
		\label{bifurca23}
\end{eqnarray}
In this case the equal angles of the kites are fixed to $60^{\circ}$. Thus for $\alpha_1=60^{\circ}$ 
the kite degenerates into an equilateral triangle while for $\alpha_1=120^{\circ}$ it becomes a 
rhombus. The equality $\eta_1(\alpha_1)=\eta_3(\alpha_1)$ gives $\alpha_1^*\simeq
75^{\circ}$. This value is rather close to $60^{\circ}$, implying that the TR phase is less stable 
with respect to deformations (within the kite geometry) of the equilateral triangle as compared 
to rhombuses or kites with $\alpha_1+\alpha_2=180^{\circ}$. In fact the difference 
$\Delta \alpha_1\equiv \left|\alpha_1^*-\alpha_1^{\rm ref}\right|$ (with 
$\alpha_1^{\rm ref}=90^{\circ}$ for rhombuses and kites with $\alpha_1+\alpha_2=180^{\circ}$, and 
$\alpha_1^{\rm ref}=60^{\circ}$ for kites with $\alpha_1+\alpha_2=240^{\circ}$), gives  
$\Delta \alpha_1\approx 15^{\circ}$, $20^{\circ}$ and $34^{\circ}$ for $240^{\circ}$-kites, 
rhombuses and $180^{\circ}$-kites, respectively.


Fig. \ref{fig4} shows the bifurcation curves for the I-N transition, $\eta_1(\alpha_1)$, and the I-T 
transition, $\eta_2(\alpha_2)$, for (a) rhombuses and (b) kites with $\alpha_1+\alpha_2=180^{\circ}$, 
as obtained from Eqns. (\ref{bifurca1}). Panel (c) corresponds to $\eta_1(\alpha_1)$ and  
$\eta_3(\alpha_1)$ (the I-TR bifurcation) for the case $\alpha_1+\alpha_2=240^{\circ}$, obtained from
Eqns. (\ref{bifurca23}). All figures also show the same bifurcation curves from
the new approach (with $r=2$). Finally in (d) the case $\alpha_1+\alpha_2=210^{\circ}$ is shown.
It is clear that the new approach gives much lower values of packing fractions at bifurcation 
than those predicted from the SPT for all the explored values of $\alpha_1$. However 
the intersections between different bifurcation curves (which can be taken to approximately 
bound the stability regions of the N, T and TR phases), are located at the same values $\alpha_1^*$. 
This result can be explained by the fact that the equality $\eta_i(\alpha_1)=\eta_j(\alpha_1)$ implies 
the same equality $\gamma_i=\gamma_j$ for both theories. 

The stability of the N, T and TR phases is bounded from below, in case of second order 
transitions, by the bifurcation curves plotted in Fig. \ref{fig4}. However, as shown in Sec. \ref{results},
the T and TR phases exhibit a transition to a N phase at high densities. Also nonuniform phases, 
not taken into account in the present study, could limit the stability of the 
orientationally ordered phases from above. To calculate the (T,TR)-N second order 
transitions, we need to perform a bifurcation analysis from T and TR phases, which we 
relegate to Sec. \ref{appendix}.

\begin{figure}
	\epsfig{file=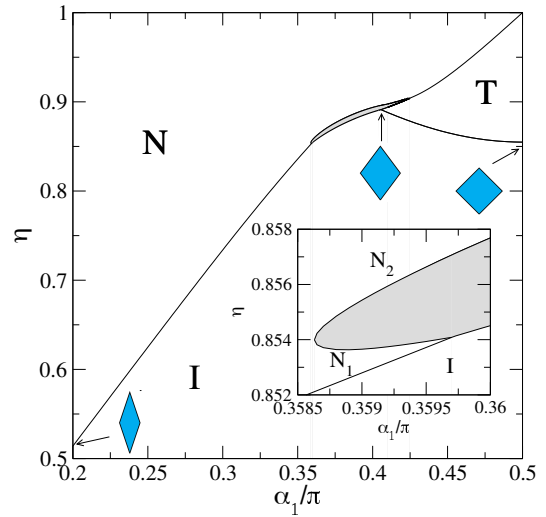,width=2.7in}
	\caption{Phase diagram of hard rhombuses ($\alpha_1=\alpha_2$) in the 
	packing fraction ($\eta$)-opening angle ($\alpha_1$) plane. The regions of stability 
	of I, T and N phases are correspondingly labeled. The  
	coexistence region of the I-N and T-N first-order phase transitions 
	(located in the neighborhood of the crossover between the I-T and T-N 
	bifurcation curves) are shaded in grey. The 
	inset shows a detail of the N$_1$-N$_2$ first-order phase transition in a small 
	$\alpha_1$ interval. Also rhombuses for different $\alpha_1$ are depicted, 
	in particular the one located at the intersection of the I-T second-order transition
	curve with the (I,TR)-binodal of the I-N (left) and TR-N (right) transitions.}
	\label{figure5}
\end{figure}

\begin{figure*}
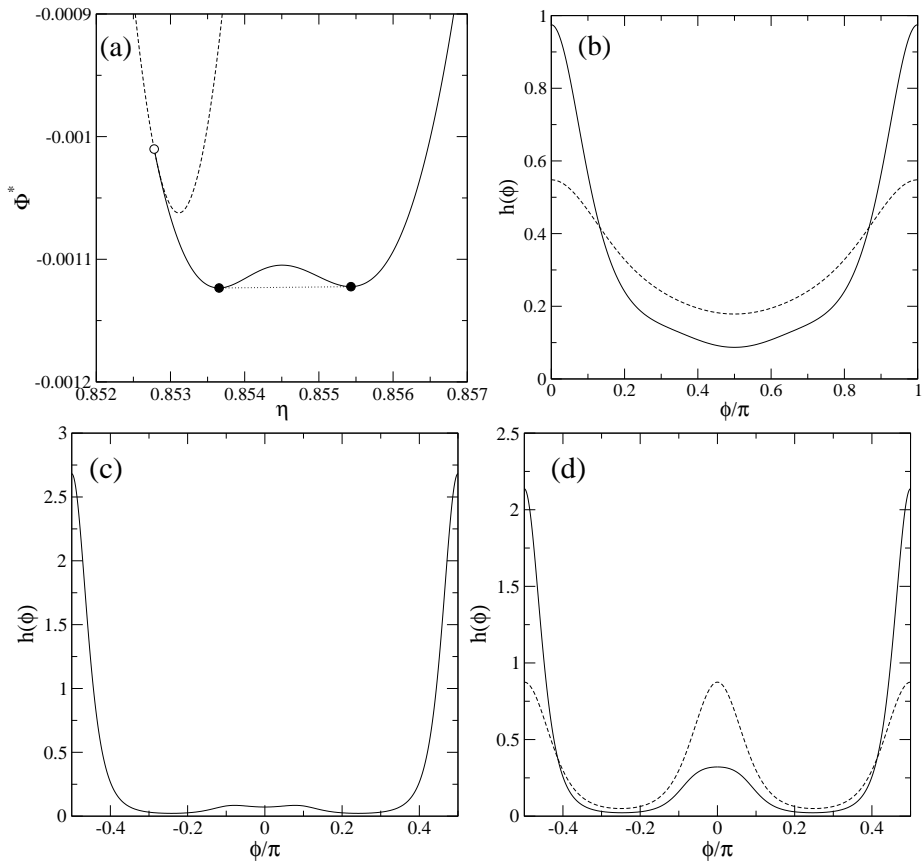

	\epsfig{file=fig6a.eps,width=2.5in}
	\epsfig{file=fig6b.eps,width=2.2in}
	\epsfig{file=fig6c.eps,width=2.3in}
	\epsfig{file=fig6d.eps,width=2.3in}
	\caption{(a) Free-energy densities of I (dashed) and N (solid) phases vs. packing fraction of hard rhombuses 
	with $\alpha_1=0.359\pi= 64.62^{\circ}$.  We have defined 
	$\Phi^*=\Phi+53.344-71.474\eta$, where a straight line has been 
	subtracted from the free-energy. 
	N-N coexistence is shown by black circles joined by a dotted line.  
	The open circle indicates the I-N bifurcation point. (b) The functions 
	$h(\phi)$ corresponding to the N$_1$ (dashed)-N$_2$ (solid) coexistence (the black circles 
	of panel (a)). (c) Orientational 
	distribution function $h(\phi)$ of the N phase coexisting with I for  
	rhombuses with $\alpha_1=0.4\pi=72^{\circ}$. (d) The functions 
	$h(\phi)$ corresponding to the T (dashed) and N (solid) phases of rhombuses with 
	$\alpha_1=0.42\pi=75.6^{\circ}$.}
	\label{fig6}
\end{figure*}

The case of kites with $\alpha_1+\alpha_2=210^{\circ}=7\pi/6$ deserves special attention. When 
$\alpha_1=30^{\circ}$ and consequently $\alpha_2=180^{\circ}$, the kite degenerates into an acute isosceles triangle, 
while for $\alpha_1=\alpha_2=105^{\circ}=7\pi/12$ particle becomes a rhombus. 
For larger values of $\alpha_1$ 
the phase diagram  is symmetric with respect to the axis $\alpha_1=105^{\circ}$. From
Fig. \ref{fig4}(d) we see that the N, T and TR phases are present in the phase diagram,
but the most striking feature is the existence of a crossover between the I-TR and I-T 
bifurcation curves. This could imply that there exist some kites which can have stable T and TR 
phases and a transition between them. In Sec. \ref{results} this case is studied in detail,  
and we will show that below this crossover the I phase exhibits a
transition to the N phase, with the latter being the stable one at high densities.

\section{Results}
\label{results}
This section is divided into three parts, each showing the phase diagrams as well as 
the orientational properties of: rhombuses (Sec. \ref{rhombus}), 
kites with the sum of the two unequal interior angles constant, $\alpha_1+\alpha_2=\text{const}$ 
(Sec. \ref{const1}), and kites with one 
of the unequal interior angles fixed to $\alpha_1=72^{\circ}$ 
(Sec. \ref{const2}).


\subsection{Hard rhombuses}
\label{rhombus}

First we calculated the phase diagram of the uniform phases of hard 
rhombuses ($\alpha_1=\alpha_2$). Apart from the I-N and I-T bifurcation curves, 
shown in Fig. \ref{fig4} (a), we also calculated the T-N bifurcation curve using 
the formalism described in Sec. \ref{appendix}. Also for those values of $\alpha_1$ 
where a first order I-N or T-N transition exists, we calculated the coexisting packing fractions 
from the equality of chemical potential and pressure of the coexisting phases. The complete 
phase diagram is shown in Fig. \ref{figure5}. We can see how the region of stability of the T phase 
is reduced as the particle shape changes from square ($\alpha_1=90^{\circ})$ 
to a critical rhombus with $\alpha_1=73^{\circ}$ (shown in the figure). The stability region of the T phase is bounded below and above by the I-N and T-N 
second-order transition curves. In the neighborhood of their intersection there exists an interval 
of $\alpha_1$ where first-order I-N and T-N transitions take place. For $\alpha_1$ 
below the intersection of the I-N bifurcation curve and the I-binodal of the I-N 
transition, there exists a N-N transition ending in a critical point.
The N-N coexistence region is shown in the inset of Fig. \ref{figure5}. 

Obviously for small values of $\alpha_1$, when the rhombus becomes highly elongated, 
the N phase is the only possible uniform phase with orientational order 
at high enough densities. This phase becomes stable at a second-order I-N transition, occurring at rather low 
packing fraction. For $\alpha_1\sim 90^{\circ}$ the T phase is the stable one at densities above a 
second-order I-T transition, at relative high packing fractions. 
As the opening angle decreases from $90^{\circ}$ and reaches a critical value 
$\alpha_1^*=73^{\circ}$,   
the T phase looses its stability. For $\alpha\gtrsim \alpha_1^*$, as the density increases, 
the T phase exhibits a transition to a N phase (see Fig. \ref{figure5}), so that 
particle axes break the fourfold symmetry and the alignment along two equivalent directors 
changes to alignment along a single director. However, as the structure of the function $h(\phi)$
indicates, this N phase keeps some tetratic correlations. 
As shown below, in the interval $[0,360^{\circ}]$ the function still exhibits four peaks 
separated by $90^{\circ}$, but two of them, 
separated by $180^{\circ}$, are much sharper and consequently the T symmetry is broken.  
The present results indicate that the second-virial DFT theories predict, for opening angles 
close to the critical value $\alpha_1^{\ast}$, the existence of first order I-N, T-N and N-N transitions, all 
of them coalescing in the same region of the phase diagram.  


The free-energy density $\Phi\equiv \beta{\cal F}a/A=\eta \varphi$ as a function of $\eta$ for 
$\alpha_1=0.359\pi=64.62^{\circ}$ is shown in Fig. \ref{fig6}(a). The free energy clearly shows the presence 
of a N-N transition. In panel (b) the coexisting orientational distribution functions 
for both uniaxial nematics for this value $\alpha_1$ are shown. 
Panel (c) shows the function $h(\phi)$ of the N phase that coexists with the I phase, for 
a value $\alpha_1=0.4\pi=72^{\circ}$ (located within the I-N first order transition region). 
We see the strong 
uniaxial ordering, with the presence of sharp peaks located at $\phi=0,180^{\circ}$, and the 
existence of small undulations around $\phi= 90^{\circ}$. Finally, in panel (d) we show $h(\phi)$
for the coexisting T and N phases at $\alpha_1=0.42\pi=75.6^{\circ}$. The former has three peaks with equal heights, 
located at $\phi=\left\{0,90^{\circ},180^{\circ}\right\}$, indicating the T symmetry 
$h(\phi)=h(\phi+\pi/2)$, while 
the latter exhibits a clear uniaxial character with the most pronounced peaks located 
at $\phi=\left\{0,180^{\circ}\right\}$, and with a secondary peak located at $\phi=90^{\circ}$, 
corresponding to the 
presence of T correlations.   

\begin{figure}
	\epsfig{file=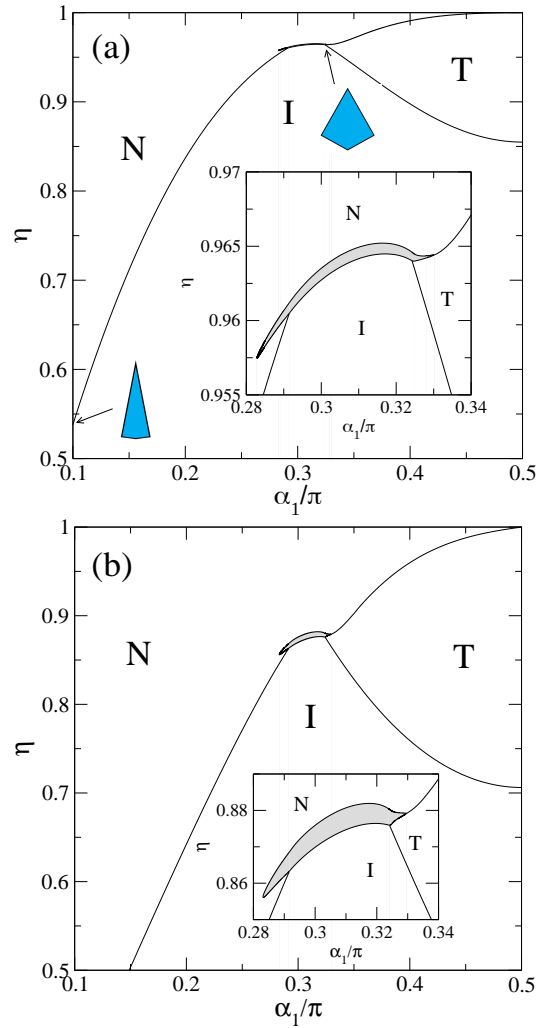,width=2.7in}
	\epsfig{file=fig7b.eps,width=2.7in}
	\caption{Phase diagrams of kites with $\alpha_1+\alpha_2=180^{\circ}$ 
	according to (a) SPT, and (b) 
	the new approach. The insets show the regions of the phase diagrams 
	where first-order transitions take place. Coexistence regions are shaded in grey.
	The stability regions of I, N and T phases are correspondingly 
	labeled. In (a) kites for two values of $\alpha_1$ are depicted.}
	\label{fig7}
\end{figure}

\subsection{Hard kites with $\alpha_1+\alpha_2=\text{const}$}
\label{const1}

The next phase diagram is that of kites with the constraint 
$\alpha_1+\alpha_2=180^{\circ}$, i.e. with the two equal angles fixed to $90^{\circ}$. 
We have used both, the 
SPT, and the new approach discussed in Sec. \ref{dft}. Results are plotted 
in Fig. \ref{fig7}(a) and (b), respectively. The fact that two of the angles of kites 
are right angles makes  
the averaged excluded area to decrease much more, as   
T ordering increases from the orientationally disordered configuration. 
If MC simulations of kites with $\alpha_1+\alpha_2=180^{\circ}$ were performed
they presumably would show a high propensity of particles to form clusters of particles joined 
by the sides adjacent to the right-angled vertexes. In turn the presence of a large amount 
of these clusters with $\alpha_1$ not acute enough is the main stabilizing mechanism for
the T phase. 
This result is confirmed in Fig. \ref{fig7}
where, according to both theories, the lower limit of stability of the T phase is reached 
for $\alpha_1\approx 58.4^{\circ}$, a critical angle significantly lower than that 
for rhombuses. In the region where the I-N, I-T and T-N 
bifurcation curves meet we again observe 
the existence of first-order phase transitions between different phases, with
the presence of a N-N transition ending in a critical point. 
Interestingly the $\alpha_1$-interval where the latter occurs is enlarged with respect to that 
of rhombuses and also takes place at higher densities. By comparing both panels we conclude that, 
within the new approach, the region of stability of the T phase is significantly enlarged, with the
second-order I-T transition occurring at lower densities. Also the I-N, T-N and N-N first-order transitions 
become stronger, with a wide density gap.

\begin{figure}
	\epsfig{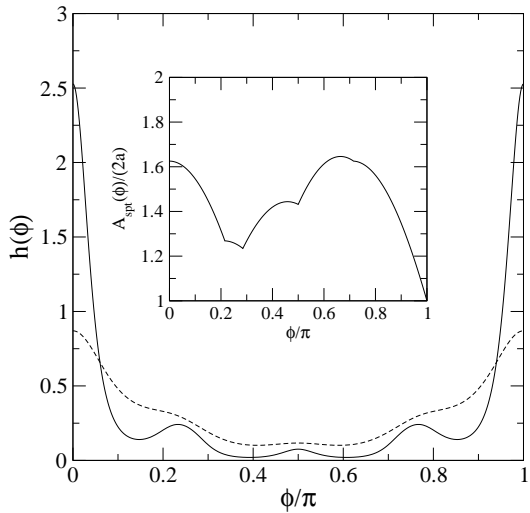}
	\caption{Orientational distribution functions, $h(\phi)$ for two coexisting 
	nematics, N$_1$ (dashed) and N$_2$ (solid) of 
	kites with $\alpha_1=0.285\pi=51.3^{\circ}$ and $\alpha_1+\alpha_2=180^{\circ}$ 
	calculated from SPT. 
	Inset: the SPT-area, $A_{\rm spt}(\phi)$, for these kites.}
	\label{fig8}
\end{figure}

In Fig. \ref{fig8} the orientational distribution functions of two coexisting 
nematics of kites with $\alpha_1=0.285\pi=51.3^{\circ}$, as calculated from SPT, are plotted. The function $h(\phi)$
for the higher-ordered nematic ($N_2$) has, apart from the main peaks 
located at $\left\{0,180^{\circ}\right\}$, three additional local maxima, whose locations are strongly 
correlated with the particle shapes. This can be seen in the inset, where we plot the 
function $A_{\rm spt}(\phi)$ for this value of $\alpha_1$. Two of the local minima 
of $A_{\rm spt}(\phi)$ are located at $\alpha_1=0.285\pi=51.3^{\circ}$ and $90^{\circ}$ (highly correlated 
with two of the positions of the local maxima of $h(\phi)$), with the other being the symmetric 
counterpart of that located at $\phi\approx 0.233\pi=41.94^{\circ}$. The latter is inside 
the interval $[0.215 \pi,0.285\pi]$, where the function $A_{\rm spt}(\phi)$ has 
a relatively low value. Thus, apart from the most favored antiparallel 
orientations of the main particle axes ($\left\{0,180^{\circ}\right\}$), 
some orientations are also favored 
to a lesser extent, due to the local minimization of the excluded area.

\begin{figure}
	\epsfig{file=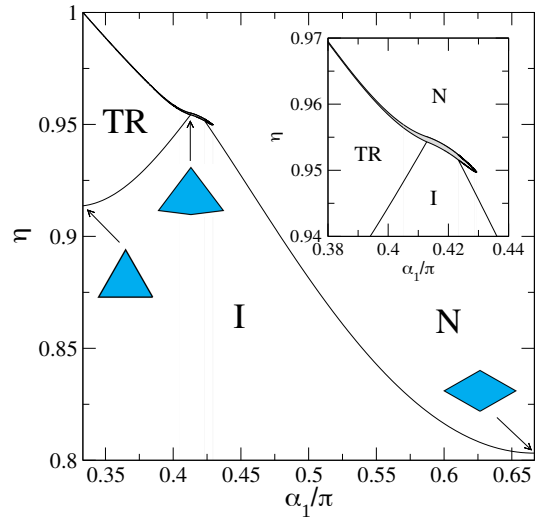,width=2.7in}
	\caption{Phase diagram of kites with $\alpha_1+\alpha_2=240^{\circ}$. The inset shows 
	a detail close to the intersection between the I-N and I-TR bifurcation curves. The 
	regions shaded in grey indicate the first order character of phase transitions. 
	Regions of stability of I, TR and N phases are correspondingly labeled. 
	Three kites with $\alpha_1=60^{\circ}$, $\alpha_1=74.3^{\circ}$ and 
	$\alpha_1=120^{\circ}$ are depicted, the 
	middle indicating the upper stability limit of TR phase.} 
	\label{fig9}
\end{figure}

We have calculated the phase diagram of kites with the constraint $\alpha_1+\alpha_2=240^{\circ}$ 
and $\alpha_1\in[60^{\circ},120^{\circ}]$ with the aim to study in what extent the TR phase, with 
the symmetry $h(\phi)=h(\phi+\pi/3)$, is still stable
by deforming an equilateral triangle within the  kite geometry. The results from the SPT 
are plotted in Fig. \ref{fig9} which shows that the region of TR phase stability, bounded 
by I-TR bifurcation curve and the TR-coexistence binodal of TR-N transition, 
ends at $\alpha_1\approx 74.3^{\circ}$ (see the shape of this kite in Fig. \ref{fig9}) 
a value not too far 
from $60^{\circ}$ indicating that the TR phase is very sensitive to these kind of deformations. 
Also, in the region where the I-TR and I-N bifurcation curves meet, the I-N transition becomes 
of first order (see the inset) which continues in a TR-N transition for lower $\alpha_1$ 
eventually keeping 
its first order character up to $\alpha_1=60^{\circ}$. We can only speculate about this fact  
close to $\eta\approx 1$ because the coexistence calculations are very difficult 
to numerically perform in this limit so we extrapolated the obtained TR and N binodals up to 
$\eta=1$.
        \begin{figure}
	\epsfig{file=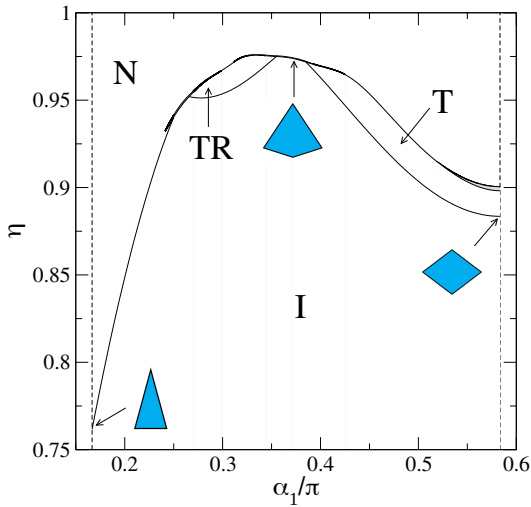,width=2.7in}
		\caption{Phase diagram of kites with $\alpha_1+\alpha_2=210^{\circ}$ following the SPT. 
		Regions of stability of I, TR and N phases are correspondingly labeled. Some 
		kites are sketched.}
		\label{fig10}
	\end{figure}

\begin{figure}
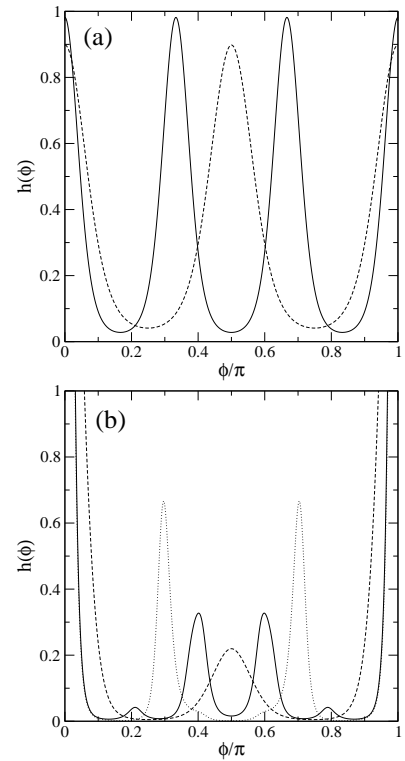

	\epsfig{file=fig11a.eps,width=2.in}
	\epsfig{file=fig11b.eps,width=2.in}
		\caption{(a) Orientational distribution functions of kites 
		with $\alpha_1+\alpha_2=210^{\circ}$ and values of the pairs 
		$(\alpha_1,\eta)=\left(54^{\circ},0.965\right)$ (solid) and 
		$\left(90^{\circ},0.922\right)$ 
		(dashed), corresponding to stable TR and T phases, respectively. 
		(b) Three different functions $h(\phi)$ corresponding to stable N 
		phases of kites with $(\alpha_1,\eta)=\left(54^{\circ},0.971\right)$ (dotted), 
		$\left(66.6^{\circ},0.975\right)$ (solid) and $\left(90^{\circ},0.935\right)$. 
		The scale of the figure 
		has been chosen so as to enhance the secondary peaks of $h(\phi)$.}
		\label{fig11}
	\end{figure}

	As we have already pointed out in Sec. \ref{bifurcation} 
	kites with $\alpha_1+\alpha_2=210^{\circ}$ 
	and $\alpha_1\in[30^{\circ},105^{\circ}]$ deserve special attention for two reasons: 
	(i) from the bifurcation analysis we showed that the T and TR phases are present in 
	the phase diagram and (ii) it is interesting to prove or discard the existence of 
	a kite with both T and TR phases and a transition between them. The complete phase 
	diagram resulting from the SPT is plotted in Fig. \ref{fig10}. 
	Indeed the TR and T phases are stable 
	and they are bounded above by a TR or T binodals of the (TR,T)-N first order phase 
	transitions except for some relatively small intervals of $\alpha_1$ where these transitions 
	becomes of second order. Two examples of equilibrium orientational distribution 
	functions $h(\phi)$ for stable T and TR phases, with their inherent symmetries
	$h(\phi+2\pi/n)$ (with $n=4$ and 6 for T and TR respectively), are shown in Fig. \ref{fig11} 
	(a). 
	As we can see from the phase diagram of Fig. \ref{fig10}, 
	for values of $\alpha_1$ close to that of the 
	intersection between I-TR and I-T bifurcation curves [see also Fig. \ref{fig4} (d)] 
	the I phase 
	exhibits a direct transition to a N phase thus discarding at all the existence 
	of a particle geometry having both stable TR and T phases. Also the packing fraction 
	values at which the TR and T phases are stable are remarkable high if we compare 
	with those of the other phase diagrams shown. Thus we expect that if we included 
	the nonuniform phases in our analysis they would be more stable than the  
	orientationally ordered uniform phases in large parts of the phase diagram.

	In Fig. \ref{fig11} (b) we plot the function $h(\phi)$ for three different stable 
	N phases for values of the opening angle of kites $\alpha_1=0.3\pi=54^{\circ}$, 
	$0.37\pi=66.6^{\circ}$ and $90^{\circ}$ and for 
	packing fractions higher than upper bounds of stability of TR, I and T phases 
	respectively (see the phase diagram of Fig. \ref{fig10}). We concentrate only on 
	the description of the secondary peaks (the much sharper main peaks are 
	located at $\{0,\pi\}$ and are outside the scale of the figure). For packing fractions 
	above the TR-phase stability region (fixing $\alpha_1=0.3\pi$) 
	the secondary peaks of the stable 
	N phase are located at $\phi\approx\{\pi/3,2\pi/3\}$ confirming the presence of important TR 
	correlations in particle orientations. As $\alpha_1$ increases up to $0.37\pi$, 
	approximately coinciding with the 
	location of the intersection between the I-TR and I-T bifurcation curves, 
	these main secondary peaks move to $\phi\approx 0.4\pi$ and $0.6\pi$,  
	approximately equal to $\alpha_1$ and its symmetric counterpart with respect to 
	$0.5\pi$. As we have already described above this issue is related with the local minima 
	of the function $A_{\rm spt}(\phi)$. It is interesting also to observe the presence 
	of two lower peaks located at $\phi\approx 0.2\pi$ and $0.8\pi$ which are also 
	related with the structure of the excluded area. Finally for $\alpha_1=0.5\pi$ 
	(the right opening angle) we observe the usual secondary peak located of 
	$\phi=\pi/2$ showing the presence of important T correlations in the fluid.
	
        \begin{figure}
	\epsfig{file=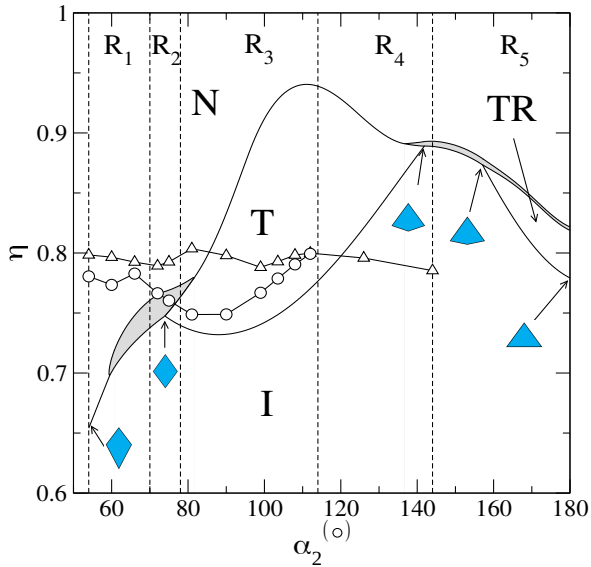,width=3.in}
		\caption{Phase diagram of hard kites with one  
		of the unequal interior angles set to 
		$\alpha_1=72^{\circ}$, while the other one is varied in 
		the interval $[54^{\circ},180^{\circ}]$. Results correspond to the new approach 
		for $\varphi_{\rm exc}[h]$, with $r=2$. The 
		regions of stability of I, N, T and TR phases are 
		correspondingly labeled. Shaded regions indicate the 
		coexistence regions of the first-order phase transitions.  
		Open circles and triangles show transitions from I 
		to liquid-crystalline uniform phases and from uniform  
		to non-uniform phases, respectively, as obtained from 
		the MC simulations of Ref. \cite{zhao1}. Different regions, 
		from R$_1$ to R$_4$, correspond to the division of the interval 
		$[54^{\circ},144^{\circ}]$ 
		introduced by the authors of Ref. \cite{zhao1}; from left to right, these regions  
		indicate the stability intervals for H$^{\rm mo}$ phase (see the text for its 
		definition), 
		an asymmetric tetratic phase T$_2$, 
		and the usual symmetric T$_1$ phase. In region R$_5$ (not calculated in 
		Ref. \cite{zhao1}), the presence of a TR phase is observed.}
		\label{fig12}
	\end{figure}

	\subsection{Hard kites with $\alpha_1=72^{\circ}$}
	\label{const2}

	Finally, we have calculated the phase diagram of kites with one of the unequal 
	opening angles fixed to $\alpha_1=72^{\circ}$, while the other one, $\alpha_2$, 
	was varied inside the interval $[54^{\circ},180^{\circ}]$. The 
	new approach for $\varphi_{\rm exc}[h]$ with $r=2$ [see Eq. (\ref{new})] was used. The 
	aim of this calculation was to compare the results obtained from 
	the implementation of our new theoretical model with recent MC simulations  
	of hard kites with the same value of $\alpha_1$ and with $\alpha_2\in[54^{\circ},144^{\circ}]$
	\cite{zhao1}. In Fig. \ref{fig12} the theoretical phase diagram, 
	together with the simulation results of Ref. \cite{zhao1}, are shown. Our 
	model predicts that,  
	as $\alpha_2$ is varied from $54^{\circ}$ to $180^{\circ}$, the I phase exhibits a sequence 
	of transitions to N ($\alpha_2\in[54^{\circ},74^{\circ}]$), T 
	($\alpha_2\in[74^{\circ},142^{\circ}]$), N again ($\alpha_2\in[142^{\circ},157^{\circ}]$), 
	and TR ($\alpha_2\in[157^{\circ},180^{\circ}]$) phases. The I-N	transitions are 
	generally of first order. 
	
	Considering only uniform phases, our analysis shows that the T and TR phases are stable up to packing fractions 
	where second- or first-order (T,TR)-N transitions occur. Fig. \ref{fig12} shows 
	the transitions from the I to liquid-crystalline phases (open circles) and from 
	these to non-uniform phases (open triangles), as obtained from the MC simulations of Ref. \cite{zhao1}.
	The authors of Ref. \cite{zhao1} divided the interval $[54^{\circ},144^{\circ}]$ in four regions 
	(enumerated in Fig. \ref{fig12} using the labels R$_i$, with $i=1,\dots,4$). 
	They claimed the existence of: (i) a molecular ordered hexatic liquid-crystal 
	phase (H$^{\rm mo}$) (in principle, this is what we call here a TR phase) in R$_1$, (ii)
	an asymmetric T phase (T$_2$) in R$_2$, (iii) the usual symmetric T phase (T$_1$) 
	in R$_3$, and (iv) a direct transition from I to non-uniform phases in R$_4$. 

	From the structure of $h(\phi)$ in the region R$_1$, with six peaks 
	separated by $60^{\circ}$ but not necessarily of the same height, the authors 
	of Ref. \cite{zhao1} concluded that H$^{\rm{mo}}$ is stable in a relatively 
	small interval of $\eta$. 
	In R$_2$ they found an angular distribution $h(\phi)$ with four peaks separated by $90^{\circ}$, 
	but these come in pairs of different height, so this was associated to an asymmetric 
	T (T$_2$) phase. Finally, in R$_3$ a distribution with nearly perfect fourfold 
	symmetry was found, which points to the usual T phase, called T$_1$.

       From the theoretical point of view, however, the definitions of the T and TR 
       phases are clearcut: 
	the symmetry $h(\phi)=h(\phi+2\pi/n)$ (with $n=4$ and $6$, respectively) must be fulfilled. 
	In case $h(\phi)\neq h(\phi+2\pi/n)$ the phase should be called N, even if  
	the secondary peaks of $h(\phi)$ (different from the main ones at $\{0,\pi\}$) 
	are sharp, pointing to important T or TR correlations in the fluid. 

\begin{figure*}
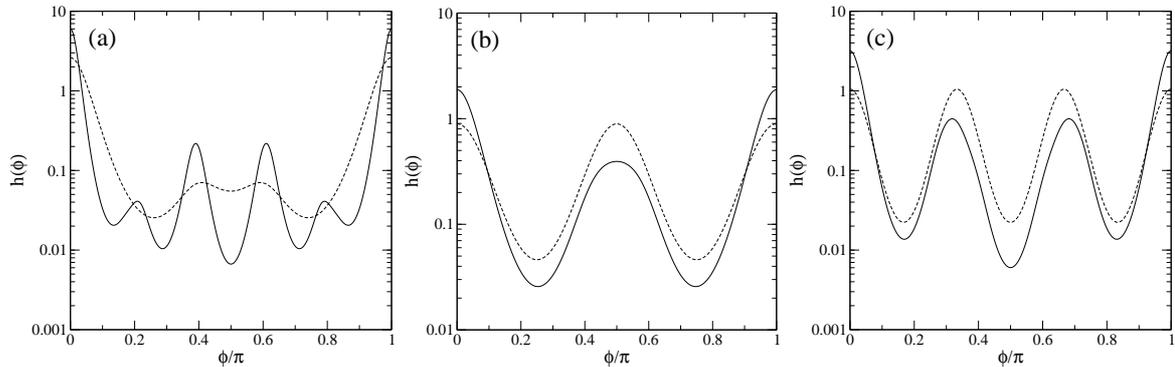

		\epsfig{file=fig13a.eps,width=2.in}
		\epsfig{file=fig13b.eps,width=1.95in}
		\epsfig{file=fig13c.eps,width=2.in}
		\caption{(a) Coexisting N distributions corresponding to the I-N 
		transitions of hard kites with $\alpha_1=72^{\circ}$, and 
		$\alpha_2=70^{\circ}$ (dashed) and 
		$150^{\circ}$ (solid). (b) Coexisting distributions corresponding to
		the T (dashed)-N (solid) transition for kites with same the $\alpha_1$ 
		and $\alpha_2=78^{\circ}$. (c) Coexisting distributions of the 
		TR (dashed)-N (solid) transition for kites with the same $\alpha_1$ and 
		$\alpha_2=171^{\circ}$.}
		\label{fig13}
	\end{figure*}

	Fig. \ref{fig13}(a) shows the function $h(\phi)$ for the coexisting N phase at 
	the I-N transition, for kites with $\alpha_2=70^{\circ}$ (just at the boundary between 
	R$_1$ and R$_2$). For values 
	of $\alpha_2$ well inside the region $R_1$, the structure of $h(\phi)$   
	is similar, except for the precise location of the secondary peaks, which change 
	with $\alpha_2$. 
	In this case, from the structure of $h(\phi)$ we can infer the existence of clear N ordering,
	with two sharp peaks at $\{0,\pi\}$, and 
	two very small secondary peaks at $\phi\approx \{0.4\pi,0.6\pi\}$, separated by a region with         a rather constant value and a weak local minimum at $\phi=0.5\pi$. 
	This approximate plateau in the interval 
	$0.4\pi$ to $0.6\pi$ indicates the existence of T correlations which, as can be seen 
	from panel (b), are much stronger in the N phase coexisting with T for  
	kites with $\alpha_2=78^{\circ}$ (a value close 
	to the boundary between the regions R$_2$ and R$_3$).  
	For $\alpha_2\in [54^{\circ},70^{\circ}]$ we only see a uniaxial N phase with  
	very small TR correlations.

	The structure of the asymmetric distribution found in R$_1$ by the simulations 
	is more similar to that we found in the N phase (coexisting with I)
	of kites with $\alpha_2=150^{\circ}$, see panel (a), or in the N phase (coexisting with TR)
	of kites with $\alpha_2=171^{\circ}$, see the panel (c); both these values are inside 
	the region R$_5$. 

	Differences in the heights of the secondary peaks of $h(\phi)$ 
	resulting from theory and simulations,	with $\alpha_2$ well inside the region R$_1$, could be explained 
	by the importance of three-body and higher correlations in the description of 
	the ordering properties of the fluid. Our theory approximates the third virial 
	coefficient of the N phase based on the second, which could explain the differences mentioned above. 
	
	Despite this, following our definitions for the orientationally ordered phases and  
	renaming H$^{\rm mo}$ to N and T$_2$ to N, the phase diagrams of MC simulations and theory are remarkable similar, 
	in particular regarding the stability of uniform orientationally-ordered phases. The I-N transition occurs
	in the regions R$_1$ and R$_2$, the I-T transition in R$_3$, and the transition from I to non-uniform phases in $R_4$,
	similar to what we found from the theoretical model (except for the presence of non-uniform phases). 
	Also the packing fractions of these	transitions are quite similar. The main drawback of the model is the 
	impossibility to study the stability of non-uniform phases, which would require a DFT for the
	one-body density profile $\rho({\bm r},\phi)$ with an accurate description of spatial correlations. An extension of the present model involving the substitution $\rho(\phi)\to\rho({\bm r},\phi)$ is simply not adequate.
	The recently developed DFT based on the Fundamental Measure Theory \cite{fmt} is expected to be a promising 
	route.

	The inclusion of non-uniform phases would probably modify the phase diagram of Fig. \ref{fig12} in the sense that
	the region where the T is now stable for $\eta\gtrsim 0.8$ would become unstable with respect to spatially ordered phases. Taking this into account we obtain a confidence interval for T-phase stability as 
	$\alpha_2\sim[74^{\circ},121^{\circ}]$, similar to that obtained from 
	simulations where the region of T$_1$-stability is $\alpha_2\sim [78^{\circ},114^{\circ}]$. 

	Finally we would like to comment on the region in the phase diagram denoted by R$_5$. 
	In this region, not simulated in Ref. \cite{zhao1}, we found that the I phase exhibits a 
	transition to a TR phase for $\alpha_2\in[157^{\circ},180^{\circ}]$, as expected 
	for kites similar to triangles and not very far from the equilateral triangle. 
	This TR phase is stable up to packing fractions where a first-order TR-N transition 
	takes place.


\section{Conclusions}
\label{conclusions}

In this paper we have presented a systematic theoretical study of the phase behavior of hard kites, with an emphasis
on the relative stability of all the possible uniform phases (I, T, TR and N). We used the SPT approximation, together
with a new approach that approximates the third virial coefficient more accurately. 
This approximation was refined by comparing the EOS of hard squares from theory and MC simulations. Several phases diagrams 
were calculated, including that of rhombuses ($\alpha_1=\alpha_2$), 
a set of them for kites with a constraint on the sum of their two unequal interior angles, 
$\alpha_1+\alpha_2=\left\{180^{\circ},240^{\circ},210^{\circ}\right\}$, and finally 
that for kites with $\alpha_1=72^{\circ}$. The latter was calculated with the aim of  
comparing with recent MC simulations \cite{zhao1}. In general we found  
first- and second-order I-(T,TR,N) and (T,TR)-N transitions, which define regions of stability 
of the uniform phases. Also we found several intervals for the opening angle where the hard-kite fluid exhibits a first-order 
N-N transition ending in a critical point.

As expected, the T phase was found to be more stable for kites with both equal angles fixed 
to $90^{\circ}$ (the constraint $\alpha_1+\alpha_2=180^{\circ}$). 
For this particular case the interval of 
$\alpha_1$ where the T is stable is the largest, $[58.4^{\circ},90^{\circ}]$; compared to that 
of rhombuses: $[73^{\circ},90^{\circ}]$. The new approach presents a stabilizing effect on the T phase, with a dramatic lowering 
of the I-T bifurcation curve, resulting in a larger T-region in the phase diagram.  
Kites with the constraint $\alpha_1+\alpha_2=240^{\circ}$ and with 
an opening angle $\alpha_1$ within the interval $\left[60^{\circ},210^{\circ}\right]$ 
(with bounds corresponding to the equilateral triangle and rhombus respectively)
have a stable TR phase for $\alpha_1\in[60^{\circ},74^{\circ}]$. We can therefore conclude 
that the TR phase is  
more sensitive to changes in particle shape (but still within the kite geometry) than
the T phase. The case $\alpha_1+\alpha_2=210^{\circ}$ is particularly interesting because 
the crossover between the I-T and I-TR bifurcation curves would suggest 
the existence of some kites exhibiting transitions between T and TR phases. However,  
we have proved this is not possible due to the presence 
of an I-N transition occurring below the crossover, the N phase being the stable one at higher densities.  
The N phase close to the crossover is peculiar, in the sense that it presents TR or T correlations 
(depending on the value of $\alpha_1$), with the orientational distribution function $h(\phi)$ 
having secondary peaks (apart from the main peaks
at $\left\{0,180^{\circ}\right\}$), located at angles $\phi$ compatible with those associated 
with the TR or T symmetries. 

By comparing the phase diagrams of kites with $\alpha_1=72^{\circ}$ obtained  
from theory and simulations, we can validate the suitability of the new approach for the prediction
of the stability of orientationally-ordered uniform phases. The interval of 
$\alpha_1$ where the T phase is stable and the densities of the I-T transition are quite similar 
in the theory and the simulations. Also similar is the structure of the orientational 
distribution function in some parts of the phase diagrams. In others this structure can be 
different, in particular regarding the relative heights of the secondary peaks, 
something that can be explained by the approximations, inherent in the theory, for the third 
and higher-order virial coefficients. 

The inclusion of non-uniform phases deserves further study. This is certainly far from trivial at the DFT level.
In this regard a DFT with an accurate description of spatial correlations would be required. An example 
of such a theory, developed for hard discorectangles and within the Fundamental 
Measure Formalism, can be found in Ref. \cite{fmt}.

\appendix

\section{Bifurcation analysis from (T,TR) phases}
\label{appendix}

The starting point in the bifurcation analysis from the T or TR phases is the nonlinear integral 
equation resulting from the equilibrium condition:
\begin{eqnarray}
	\frac{\delta \varphi [h]}{\delta  h(\phi)}=\lambda\Rightarrow 
	h(\phi)=\exp{\left[\lambda-\frac{\delta \varphi_{\rm exc}[h]}{\delta h(\phi)}\right]},
	\label{primera}
\end{eqnarray}
where $\lambda$ is a Lagrange multiplier that guarantees the normalization $\int_0^{\pi}d\phi h(\phi)=1$. 
Taking into account Eqn. (\ref{new}), we have 
\begin{eqnarray}
	&&\frac{\delta \varphi_{\rm exc}[h]}{\delta h(\phi)}=\psi\left[h;\eta\right]
	\frac{\delta \gamma[h]}{\delta h(\phi)},\nonumber\\
	&&\frac{\delta \gamma[h]}{\delta h(\phi)}=\int_0^{\pi}d\phi' h(\phi+\phi')\frac{A_{\rm spt}(\phi')}{a}
	\nonumber\\&&\hspace{3cm}=2\sum_{k\geq 0} \gamma_k h_k\cos(2k\phi),
	\label{segunda}
\end{eqnarray}
where we have used the Fourier representation (\ref{fourier}) of $h(\phi)$ and the definition 
(\ref{coeff}) for the coefficients $\gamma_k$. Also the shorthand notation 
\begin{eqnarray}
	&&\psi\left[h;\eta\right]=
	\frac{\eta}{1-\eta}+\left(2\gamma[h]-1\right)
	\nonumber\\
	&&\times \left(\frac{1}{2}+r\eta\right)\left(\frac{\eta}{1-\eta}+\log(1-\eta)\right).
	\label{psi}
\end{eqnarray}
has been used. From Eqs. (\ref{primera}) and (\ref{segunda}) we obtain 
\begin{eqnarray}
h(\phi)=e^{\lambda}
	\exp{\left\{-2\psi\left[h;\eta\right]
	\sum_{k\geq 1} \gamma_kh_k\cos(2k\phi)\right\}},
	\label{inter}
\end{eqnarray}
where $\lambda$ can be calculated from 
\begin{eqnarray}
	e^{-\lambda}=\int_0^{\pi} d\phi' \exp{\left\{-2\psi\left[h;\eta\right]
        \sum_{k\geq 1} \gamma_kh_k\cos(2k\phi')\right\}},
\end{eqnarray}
which obviously guarantees the normalization condition. 
Multiplying (\ref{inter}) by $\cos(2j\phi)$, integrating over $\phi$, and using again 
the expansion (\ref{fourier}), we obtain 
\begin{eqnarray}
	&&h_j =2e^{\lambda}\int_0^{\pi}d\phi\cos(2j\phi)\nonumber\\
	&&\times \exp{\left\{-2\psi\left[h;\eta\right]
        \sum_{k\geq 1} \gamma_kh_k\cos(2k\phi)\right\}}.
	\label{casi}
\end{eqnarray}
Now a small perturbation of the T phase is introduced, resulting in a N phase with orientation 
distribution function
\begin{eqnarray}
	&&h_{\rm N}(\phi)=h_{\rm T}(\phi)+\frac{1}{\pi}
	\sum_{j\geq 1} h_{2j-1}\cos[2(2j-1)\phi], \nonumber\\ 
	&&h_{\rm T}(\phi)=\frac{1}{\pi}\left(1+\sum_{j\geq 1} h_{2j} \cos(4j\phi)\right),
\end{eqnarray}
with $h_{2j-1}\ll h_{2j}$. We define the quantity 
\begin{eqnarray}
	T(\phi)\equiv \exp{\left\{-\psi_{\rm T}\left[h;\eta\right]
	\sum_{k\geq 1} \gamma_{2k}h_{2k}\cos(4k\phi)\right\}},
\end{eqnarray}
with $\psi_{\rm T}\left[h;\eta\right]$ calculated from Eq. (\ref{psi}) with the anisometry parameter 
having a T symmetry:
\begin{eqnarray}
	\gamma_{\rm T}[h]=\gamma_0+\frac{1}{2}\sum_{k\geq 1} \gamma_{2k} h_{2k}^2.
\end{eqnarray}
Expanding Eqn. (\ref{casi}) for $j=2n-1$ up to first order in $\{h_{2k-1}\}$, 
and using the symmetry of the T phase (implying $\int_0^{\pi}d\phi T(\phi)\cos\left[2(2j-1)\phi\right]=0$), 
we obtain 
\begin{eqnarray}
	&&h_{2n-1}=-\frac{2\psi_0(\eta)}{\int_0^{\pi} 
	d\phi' T(\phi')}\int_0^{\pi} d\phi T(\phi)\sum_{k\geq 1}\gamma_{2k-1} h_{2k-1}\nonumber\\
	&&\times 
	\left\{\cos[4(k+n-1)\phi]+\cos[4(k-n)\phi]\right\}\\
	&&\Rightarrow h_{2n-1}=-\psi_0(\eta)\sum_{k\geq 1} \gamma_{2k-1} \nonumber\\
	&&\times \left[
		h_{2(k+n-1)}+h_{2|k-n|}\right] h_{2k-1}.
	\label{final0}
\end{eqnarray}
Here we have used the definition 
$h_{2k}=2\int_0^{\pi} d\phi T(\phi)\cos(4k\phi)/\int_0^{\pi} d\phi' T(\phi')$ while the 
function $\psi_0(\eta)$ is the same as (\ref{psi}) with the 
substitution $\gamma[h]\to\gamma_0$.   

Defining now the column vector ${\bm c}$ with coordinates $c_k=h_{2k-1}$, $k=1,\dots,m/2$ (with $m$ an even number) 
and the matrix $B$ with elements 
\begin{eqnarray}
	&&b_{nk}=\delta_{nk}+\psi_0(\eta)\gamma_{2k-1}\left(h_{2(k+n-1)}+h_{2|k-n|}\right),\nonumber\\
	&&\hspace{2cm}n,k=1,\dots,\frac{m}{2}, 
\end{eqnarray}
the Eqn. (\ref{final0}) can be put in the matrix form $B{\bm c}=\boldsymbol{0}$ which has a nontrivial 
solution for ${\bm c}$ if and only if
\begin{eqnarray}
	{\cal B}(\eta,\{h_{2k}\})\equiv \text{det}\left(B\right)=0. \label{final}
\end{eqnarray}
$m/2$ is the total number of even Fourier amplitudes $\{h_1,h_3,\dots,h_{m-1}\}$, which  
are of same order, say $\sim \epsilon$, in the perturbative expansion of $h(\phi)$ around 
$h_{\rm T}(\phi)$. We need to take $h_{2(k+n-1)}=0$ if $k+n-1>m/2$ and $h_{2|k-n|}=2$ for $k=n$. 

We solve Eqn. (\ref{final}) iteratively for the present as well as for the SPT approach (obtained 
by replacing $2\gamma_0-1$  by $0$ in Eqn. (\ref{psi})) to find 
the T-N bifurcation value of $\eta$, once the equilibrium Fourier amplitudes of the T phase 
$\{h_{2k}\}$ have been obtained (these in turn depend on $\eta$). In most of the calculated T-N bifurcations we found 
that assuming all even Fourier amplitudes $\{h_{2k-1}\}$ to have the same order $\epsilon$ exactly gives a value $\eta^*$ in
agreement with that found from the free-energy minimization with respect to all $\{h_j\}$ (odd and even) for a given $\eta$ 
(and extrapolating $\eta\to \eta^*$, which gives $h_{2k-1}\to 0$).

The bifurcation analysis can also be implemented for a small perturbation of the TR phase, resulting in a N phase with
\begin{eqnarray}
	&&\hspace{-1.3cm}h_{\rm N}(\phi)=h_{\rm TR}(\phi)+
	\frac{1}{\pi}\sum_{i=1,2}\sum_{j\geq 1}h_{3j-i}\cos\left[2(3j-i)\phi\right],
	\\ 
	&&\hspace{-1.3cm}h_{\rm TR}(\phi)=\frac{1}{\pi}
	\left(1+\sum_{j\geq 1} h_{3j}\cos(6j\phi)\right), 
	h_{3j-i}\ll h_{3j}.
\end{eqnarray}
This analysis can be realized using the same procedure as for the bifurcation from the T phase. The result is: 
\begin{eqnarray}
	&&h_{3n-l}=-\frac{2\psi_0(\eta)}{\int_0^{\pi} d\phi'T(\phi')}
	\sum_i\sum_j \gamma_{3j-i}h_{3j-i}
	\int_0^{\pi} d\phi T(\phi)\nonumber\\
	&&\times \left\{\cos[6(n+j-1)\phi)\delta_{l+i,3}
	+\cos[6(n-j)\phi]\delta_{l-i,0}\right\}\nonumber\\
	&&\\
	&&\Rightarrow h_{3n-l}=-\psi_0(\eta)\sum_i\sum_j \gamma_{3j-i}\nonumber\\ 
	&&\times \left[h_{3(n+j-1)}\delta_{l+i,3}+h_{3|n-j|}\delta_{l-i,0}\right]
	h_{3j-i},
\end{eqnarray}
where in this case 
\begin{eqnarray}
	&&T(\phi)=\exp{\left\{-2\psi_{\rm TR}\left[h;\eta\right]
	\sum_{k\geq 1}\gamma_{3k}h_{3k}\cos(6k\phi)\right\}},\nonumber\\
	&&\\
	&&\gamma_{\rm TR}[h]=\gamma_0+\frac{1}{2}\sum_{k\geq 1}
	\gamma_{3k}h_{3k}^2.
\end{eqnarray}

Defining the vector 
${\bm c}=\left({\bm c}^{(1)},{\bm c}^{(2)}\right)^T$ with 
coordinates $c^{(i)}_k=h_{3k-i}$ ($i=1,2$), $k=1,\dots, m/3$ (with $m$ a multiple of 3), and the matrix 
\begin{eqnarray}
	B\equiv \begin{pmatrix}
		B^{(1,1)} & B^{(1,2)}\\
		B^{(2,1)} & B^{(2,2)}
	\end{pmatrix}
\end{eqnarray}
with matrix elements 
\begin{eqnarray}
	&&b_{nj}^{(l,i)}=\delta_{n,j}\delta_{l,i}+\psi_0(\eta)\gamma_{3j-i}\nonumber\\&&\hspace{1cm}\times 
	\left[h_{3(n+j-1)}\delta_{l+i,3}+h_{3|n-j|}\delta_{l-i,0}
	\right],\nonumber\\&& n,j=1,\dots,\frac{m}{3},
\end{eqnarray}
we solve Eqn. (\ref{final}) to find the packing fraction at bifurcation. Again we take 
$h_{3(n+j-1)}=0$ if $n+j-1>m/3$ and $h_{3|n-j|}=2$ if $n=j$.

\acknowledgements

Financial support under grant FIS2017-86007-C3-1-P
from Ministerio de Econom\'{\i}a, Industria y Competitividad (MINECO) of Spain,
and PGC2018-096606-B-I00 from Agencia Estatal de Investigaci\'on-Ministerio de Ciencia e Innovaci\'on of Spain,
is acknowledged.


\begin{thebibliography}{37}
\bibitem{onsager} L. Onsager, Ann. N. Y. Acad. Sci. {\bf 51}, 627 (1949).
\bibitem{frenkel} D. Frenkel, H. N. W. Lekkerkerker, and A. Stroobants, Nature {\bf 332}, 
	822 (1988).
\bibitem{veerman} J. A. C. Veerman and D. Frenkel, Phys. Rev. A {\bf 45}, 5632 (1992).
\bibitem{allen} A. Samborski, G. T. Evans, C. P. Mason, and M. Allen, 
	Mol. Phys. {\bf 81}, 263 (1994).
\bibitem{jackson} S. C. McGrother, D. C. Willianson and G. Jackson, J. Chem. Phys. 
	{\bf 104}, 6755 (1996).
\bibitem{bolhuis} P. Bolhuis and D. Frenkel, J. Chem. Phys. {\bf 106}, 666 (1997).
\bibitem{mulder} P. I. Teixeira, A. J. Masters, and B. M. Mulder, Mol. Cryst. Liq. 
	Cryst. {\bf 323}, 167 (1998).
\bibitem{kooij}	F. M. van der Kooij, K. Kassapidou, and H. N. W. Lekkerkerker, 
	Nature {\bf 406}, 868 (2000).
\bibitem{wensink} H. H. Wensink and H. N. W. Lekkerkerker, Mol. Phys. {\bf 107}, 2111 (2009).
\bibitem{cheng} D. Sun, H.-J. Sue, Z. Cheng, Y. Mart\'{\i}nez-Rat\'on and E. Velasco, 
	Phys. Rev. E {\bf 80}, 041704 (2009).
\bibitem{schiling} P. Pfleiderer and T. Schilling, Phys. Rev. E {\bf 75}, 
	020402(R) (2007).
\bibitem{cinacchi} G. Cinacchi and J. Duijneveldt, J. Phys. Chem. Lett. 
	{\bf 1}, 787 (2010).
\bibitem{odriozola} G. Odriozola, J. Chem. Phys. {\bf 136}, 134505 (20012).
\bibitem{yang} Y. Yang, G. Chen, S. Thanneeru, J. He, K. Liu, and Z. Nie, Nature 
	Communications {\bf 9}, 4513 (2018).
\bibitem{cuetos} A. Cuetos, M. Denninson, A. Masters and A. Patti, Soft Matter {\bf 13}, 
	4720 (2017).
\bibitem{dijkstra2} S. Dussi, N. Tasios, T. Drwenski, R. van Roij, and M. Dijkstra, Phys. 
	Rev. Lett. {\bf 120}, 177801 (2018).
\bibitem{dijkstra3} M. Chiappini, T. Drwenski, R. van Roij, and M. Dijkstra, Phys. Rev. Lett. 
	{\bf 123}, 068001 (2019).
\bibitem{rafael} E. M. Rafael, D. Corbett, A. Cuetos, and A. Patti, 
	Soft Matter {\bf 16}, 5565 (2020).
\bibitem{mederos} L. Mederos, E. Velasco, and Y. Mart\'{\i}nez-Rat\'on, J. Phys.: Condens. 
	Matter {\bf 26}, 463101 (2014).
\bibitem{chaikin} K. Zhao, C. Harrison, D. Huse, W. B. Russel, and P. M. Chaikin
	Phys. Rev. E {\bf 76}, 040401(R) (2007).
\bibitem{zhao3} K. Zhao, R. Bruinsma, and T. G. Mason, PNAS {\bf 108}, 2684 (2011).
\bibitem{zhao5} K. Zhao, R. Bruinsma, and T. G. Mason, Nature  Communications {\bf 3}, 801 (2012).	
\bibitem{frenkel1} K. W. Wojciechowski and D. Frenkel, Comp. Meth. Sci. Tech. {\bf 10}, 235 (2004).
\bibitem{donev} A. Donev, J. Burton, F. H. Stillinger, and S. Torquato, Phys. Rev. B {\bf 73}, 054109 (2006).
\bibitem{schlacken} H. Schlacken, H.-J. Mogel, and P. Schiller, Mol. Phys. {\bf 93}, 777 (1998).
\bibitem{MR3}Y. Mart\'{\i}nez-Rat\'on, E. Velasco, and L. Mederos, J. Chem. Phys. {\bf 122}, 064903 (2005).
\bibitem{MR4}Y. Mart\'{\i}nez-Rat\'on, E. Velasco, and L. Mederos, J. Chem. Phys. {\bf 125}, 014501 (2006).
\bibitem{dijkstra}A. P. Gantapara, W. Qi, and M. Dijkstra, Soft Matter {\bf 11}, 8684 (2015).
\bibitem{MR2} Y. Mart\'{\i}nez-Rat\'on, A. D\'{\i}az-De Armas and E. Velasco, 
	Phys. Rev. E {\bf 97}, 052703 (2018).

\bibitem{glotzer} J. A. Anderson, J. Antonaglia, J. A. Millan, M. Engel, and S. C. Glotzer, Phys. 
	Rev. X {\bf 7}, 021001 (2017).
\bibitem{escobedo} C. Avenda\~no and F. A. Escobedo, Soft Matter {\bf 8}, 4675 (2012).
\bibitem{zhao4} Z. L. Hou, K. Zhao, Y. W. Zong, and T. G. Mason, Phys. Rev. Matt. 
	{\bf 3}, 015601 (2019).
\bibitem{szabi}S. Mizani, P. Gurin, R. Aliabadi, H. Salehi, and S. Varga, 
	J. Chem. Phys. {\bf 153}, 034501 (2020).
\bibitem{mayoral} K. Mayoral and T. G.  Mason, Soft Matter {\bf 10}, 4471 (2014).
 \bibitem{zhao6}K. Zhao and  T. G. Mason, J. Phys.: Condens. Matter {\bf 26}, 152101 (2014).
	\bibitem{zhao7} P.-Y. Wang and T. G. Mason, Nature {\bf 561}, 94 (2018).
\bibitem{zhao1} Z. Hou, Y. Zong, Z. Sun, F. Ye, T. G. Mason, and 
	K. Zhao, Nature Commun. {\bf 11}, 2064 (2020). 
\bibitem{boublik} T. Boublik and I. Nezbeda, Collect. Czech. 
	Commun. {\bf 51}, 2301 (1986).
\bibitem{tarjus} G. Tarjus, P. Viot, S. M. Ricci, and J. Talbot, 
	Mol. Phys. {\bf 73}, 773 (1991).
\bibitem{isihara}A. Isihara, J. Chem. Phys. {\bf 18}, 1446 (1950).
\bibitem{kihara} T. Kihara, Rev. Mod. Phys. {\bf 25}, 831 (1953).
\bibitem{fmt}R. Wittmann, C. E. Sitta, F. Smallenburg, and H. L\"owen, J. Chem. Phys. {\bf 147}, 134908 (2017).
\end{thebibliography}
\end{document}